\documentclass[12pt]{iopart}
\usepackage{iopams}
\usepackage{cite}
\usepackage{array}
\usepackage[dvipsnames]{xcolor}
\usepackage{graphicx}
\usepackage[colorlinks=true,citecolor=blue]{hyperref}
\usepackage{ulem}
\usepackage{soul}
\setstcolor{Red}
\usepackage{cancel}
\newcolumntype{L}{>{$}l<{$}} 
\eqnobysec
\def\sfrac#1#2{\textstyle{\frac{#1}{#2}}}

\begin{document}

\title[Vacuum polarization on topological black holes]{Vacuum polarization on topological black holes}

\author{Thomas Morley}

\address{Consortium for Fundamental Physics, School of Mathematics and Statistics,\\ Hicks Building, Hounsfield Road, Sheffield. S3 7RH United Kingdom}
\ead{TMMorley1@sheffield.ac.uk}

\author{Peter Taylor}
\address{Centre for Astrophysics and Relativity, School of Mathematical Sciences,\\ Dublin City University, Glasnevin, Dublin 9, Ireland}
\ead{Peter.Taylor@dcu.ie}

\author{Elizabeth Winstanley}
\address{Consortium for Fundamental Physics, School of Mathematics and Statistics,\\ Hicks Building, Hounsfield Road, Sheffield. S3 7RH United Kingdom}
\ead{E.Winstanley@sheffield.ac.uk}

\begin{abstract}
We investigate quantum effects on topological black hole space-times within the framework of quantum field theory on curved space-times. Considering a quantum scalar field, we extend a recent mode-sum regularization prescription for the computation of the renormalized vacuum polarization to asymptotically anti-de Sitter black holes with nonspherical event horizon topology. In particular, we calculate the vacuum polarization for a massless, conformally-coupled scalar field on a four-dimensional topological Schwarzschild-anti-de Sitter black hole background, comparing our results with those for a spherically-symmetric black hole. 
\end{abstract}

\pacs{04.62.+v, 04.70.Dy}

\vspace{2pc}
\noindent{\it Keywords}: vacuum polarization, topological black holes

\section{Introduction}
\label{sec:intro}

In quantum field theory on curved space-time (QFTCS), the renormalized stress-energy tensor (RSET) $\langle {\hat {T}}_{\mu \nu} \rangle $ is an object of central importance. Via the semi-classical Einstein equations 
\begin{equation}
G_{\mu \nu }+ \Lambda g_{\mu \nu } = 8\pi \langle {\hat {T}}_{\mu \nu} \rangle ,
\label{eq:scee}
\end{equation}
(where $G_{\mu \nu }$ is the Einstein tensor, $\Lambda $ the cosmological constant and $g_{\mu \nu }$ the metric tensor)
the RSET governs the back-reaction of the quantum field on the space-time geometry.
The stress-energy tensor operator ${\hat {T}}_{\mu \nu }$ involves products of field operators evaluated at the same space-time point and is therefore divergent.
In the point-splitting approach pioneered by DeWitt and Christensen \cite{DeWitt:1975ys,Christensen:1976vb,Christensen:1978yd}, this divergence is regularized by considering the operators acting on two closely-separated space-time points.  
The RSET is computed by subtracting off the divergences, which are purely geometrical in nature and independent of the quantum state under consideration. 
The parametrix encoding these geometric divergent terms was originally constructed using a DeWitt-Schwinger expansion \cite{DeWitt:1975ys,Christensen:1976vb,Christensen:1978yd}. This prescription was made more precise by Wald \cite{Wald:1977up,Wald:1995yp} who gave a set of axioms that (almost) uniquely determine the RSET. Wald further showed that encoding the divergences using Hadamard elementary solutions, of which the DeWitt-Schwinger expansion can be thought of as a special case, produced an RSET satisfying these axioms. The Hadamard prescription provides a more elegant and  general prescription than the DeWitt-Schwinger approach in that it can be applied for arbitrary field mass and arbitrary dimensions (see, for example, \cite{Decanini:2005eg}).

In practice, the computation of the RSET on space-times other than those with maximal symmetry (see, for example, \cite{Kent:2014nya,Ottewill:2012mq}) is a challenging task. Of particular interest are black hole space-times, and there is a long history of RSET computations on asymptotically flat Schwarzschild black hole backgrounds, for both quantum scalar fields \cite{Fawcett:1983dk,Howard:1984qp,Howard:1985yg,Jensen:1992mv,Ottewill:2010hr} and fields of higher spin \cite{Elster:1984hu,Groves:2002mh,Jensen:1988rh,Jensen:1995qv}.
In four space-time dimensions, Anderson, Hiscock and Samuel (AHS) \cite{Anderson:1993if,Anderson:1994hg} have developed a general methodology for finding the RSET on a static, spherically-symmetric black hole. 
Their method makes heavy use of WKB approximations and the RSET is given as a sum of two parts, the first of which is analytic and the second of which requires numerical computation.
More recently, Levi, Ori and collaborators \cite{Levi:2016quh,Levi:2016exv} have developed a new method (dubbed ``pragmatic mode-sum regularization'' \cite{Levi:2015eea}) for finding the RSET which does not rely on WKB approximations and
has the advantage that it can be applied to stationary as well as static black holes \cite{Levi:2016exv}.

Given the challenges of computing the RSET, it is instructive to consider instead the vacuum polarization (VP) $\langle {\hat {\phi }}^{2} \rangle $ of a quantum scalar field ${\hat {\phi }}$. 
Unlike the tensor RSET, the VP is a scalar object and hence cannot distinguish between the future and past event horizons of a black hole.
Nonetheless, the VP shares many features with the RSET, for example, if the VP diverges on a horizon, then it is likely that the RSET will also diverge there.
The VP has been computed on asymptotically flat, spherically-symmetric, four-dimensional black hole space-times (see, for example, \cite{Candelas:1980zt,Candelas:1984pg,Anderson:1989vg,Breen:2015hwa,Levi:2015eea,Levi:2016esr})
using both the AHS and Levi-Ori methods. 
The AHS method uses a Euclideanized black hole space-time, and is therefore most amenable for the computation of the VP in the Hartle-Hawking state \cite{Hartle:1976tp}, while the Levi-Ori method employs a Lorentzian metric and has been applied to the Boulware \cite{Boulware:1975pe} and Unruh \cite{Unruh:1976db} states.
The AHS method has the disadvantage that, unless the field is massless and conformally coupled, the analytic part of the expression for the VP diverges at the black hole event horizon.
The numeric part also diverges at the horizon, resulting in a quantity which is finite overall \cite{Winstanley:2007tf}. 
Using Green-Liouville asymptotics, Breen and Ottewill \cite{Breen:2010ux} improved the AHS method by writing the VP as a sum of terms, each of which is manifestly finite. Their method also extends to the RSET \cite{Breen:2011af,Breen:2011aa}. Another approach is to match the Hadamard parametrix to a mode-sum in Minkowski space-time in a suitable choice of coordinates \cite{Ottewill:2010bq,Ferreira:2014ina}, but this approach is not very general and requires a bespoke mapping between Minkowski space-time and the black hole being considered. However, this matching method has been successfully applied to nonspherically-symmetric space-times where the AHS method is not applicable  \cite{Ottewill:2010bq,Ferreira:2014ina}.

With the exception of \cite{Ferreira:2014ina}, the above discussion focuses on asymptotically flat black hole geometries. 
Asymptotically anti-de Sitter (adS) black holes have received rather less attention in the QFTCS literature, despite their inherent interest due to the adS/CFT (conformal field theory) correspondence (see, for example, \cite{Aharony:1999ti} for a review).
The VP has been computed for a massless, conformally-coupled scalar field on a four-dimensional spherically-symmetric Schwarzschild-adS black hole using the AHS method \cite{Flachi:2008sr} and also on a static, spherically-symmetric adS black hole with asymptotically Lifschitz geometry \cite{Quinta:2016eql}.
An AHS-like method has also been used to find the VP on an asymptotically adS black hole with cylindrical rather than spherical symmetry 
\cite{DeBenedictis:1998be}.
The corresponding calculations on the three-dimensional, asymptotically adS BTZ black hole \cite{Banados:1992wn,Banados:1992gq,Carlip:1995qv}
are rather simpler than those for four-dimensional black holes and both the VP and RSET for a conformally-coupled scalar field can be found in closed form (both when there is a black hole event horizon and in the naked singularity case \cite{Steif:1993zv,Lifschytz:1993eb,Shiraishi:1993nu,Shiraishi:1993ti,Binosi:1998yu,Worden:2018}).
These closed-form expressions have been used to study the back-reaction of the quantum field on the space-time geometry via the semi-classical Einstein equations (\ref{eq:scee}) \cite{Martinez:1996uv,Casals:2016ioo,Casals:2016odj}. 

In asymptotically adS space-time, unlike asymptotically flat space-time, black holes do not necessarily have spherical event horizon topology (see, for example, \cite{Birmingham:1998nr,Brill:1997mf,Lemos:1994fn,Lemos:1994xp,Lemos:1995cm,Vanzo:1997gw,Cai:1996eg,Mann:1996gj,Smith:1997wx,Mann:1997zn}). 
This context has received little attention in the literature on semi-classical effects, but provides an interesting dichotomy between the classical and semi-classical Einstein equations. For example, given an asymptotically adS black hole with an event horizon whose geometry is flat, it is straightforward to make identifications that produce a black cylinder or a black torus.  However, as these identifications correspond to a choice of boundary conditions, the classical equations themselves are not sensitive to this choice since the PDEs are quasi-local in nature (though obviously a particular solution is picked out by the choice of boundary conditions). 
On the other hand, the semi-classical Einstein equations (\ref{eq:scee}) are sourced by the expectation value of a
field operator in a given quantum state, a state which requires global information in order to be defined and which is sensitive to any such identifications that distinguish between, say, a cylinder and a torus.
Since the quantum stress-energy tensors on backgrounds with different topologies (but with the
same geometry) are different, the back-reaction effects will be very much sensitive to the global topology. 
Though we do not consider different identifications for the same black hole geometry in this paper, this context does provide an additional motivation and as a first step in these directions, it is necessary to consider quantum fields on
the backgrounds of asymptotically adS black holes with horizons of nonspherical topology. More specifically, we calculate the VP for a massless, conformally-coupled quantum scalar field on the backgrounds of Schwarzschild-adS black holes with event horizons of spherical, planar and hyberboloidal topology.

We work in the Hartle-Hawking state \cite{Hartle:1976tp} and follow the recent ``extended coordinates'' methodology of \cite{Taylor:2016edd,Taylor:2017sux},
which gives an efficient numerical method for finding the VP on the Euclideanized space-time. 
In \cite{Taylor:2016edd,Taylor:2017sux}, the extended coordinates method is developed for static, spherically-symmetric black holes in four or more space-time dimensions.
Here we apply the extended coordinates approach to black holes with nonspherical event horizon topology.
For our purposes, a distinct advantage of the extended coordinates approach is that the asymptotically adS nature of the space-times we consider does not present any particular difficulties. 

The outline of the paper is as follows.
We review the geometry of topological black holes in Section \ref{sec:bh}, before deriving an expression for the Euclidean Green's function for a massless, conformally-coupled, scalar field on these backgrounds in Section \ref{sec:scalar}.
The renormalized VP will be calculated using Hadamard renormalization, and in Section \ref{sec:hadamard} we introduce the Hadamard parametrix, employing the extended coordinates method of \cite{Taylor:2016edd,Taylor:2017sux} to write this as a mode sum, which is amenable to numerical evaluation.
Our numerical methodology is outlined in Section \ref{sec:renVP}, together with our numerical results.
Section \ref{sec:conc} contains our conclusions and further discussion.
Throughout this paper, we use mostly plus space-time conventions, and units in which $G=\hbar = c = k_{B}=1$.

\section{Topological black holes}
\label{sec:bh}

We consider static black hole solutions of the vacuum Einstein equations with a negative cosmological constant $\Lambda $:
\begin{equation}
G_{\mu \nu }+ \Lambda g_{\mu \nu } = 0.
\end{equation}
The event horizon is a two-surface of constant curvature, and the metric takes the following form, in Schwarzschild-like coordinates
\cite{Birmingham:1998nr,Brill:1997mf,Lemos:1994fn,Lemos:1994xp,Lemos:1995cm,Vanzo:1997gw,Cai:1996eg,Mann:1996gj,Smith:1997wx,Mann:1997zn}:
\begin{equation}
\label{eq:metric}
\rmd s^{2}=-f(r)\, \rmd t^{2}+\frac{\rmd r^{2}}{f(r)}+r^{2} \,\rmd \Omega_{k}^{2},
\end{equation}
where $k\in \{ -1,0,1 \} $, corresponding to negative, zero and positive horizon curvature respectively.
The function $f(r)$ is given by 
\begin{equation}
f(r)=k-\frac{2M}{r}+\frac{r^{2}}{L^{2}},
\label{eq:f}
\end{equation}
where $M$ is the black hole mass and $L={\sqrt {-3/\Lambda }}$ is the adS curvature length-scale.
The two-metric $\rmd \Omega _{k}^{2}$ is 
\begin{equation}
\rmd \Omega_{k}^{2}=\rmd \theta^{2}+\mathcal{F}_{k}^{2}(\theta) \, \rmd \varphi^{2},
\label{eq:omegak}
\end{equation}
with
\begin{equation}
\mathcal{F}_{k}(\theta)=\left\{\begin{array}{ll}
\sin\theta, & k=1,\\
\theta, & k=0,\\
\sinh\theta, & k=-1.
\end{array}\right.
\label{eq:Fk}
\end{equation}
For all values of $k$, the coordinate $\varphi $ takes values in the interval $[0,2\pi )$ and the event horizon is located at $r=r_{\mathrm{h}}$, with $f(r_{\mathrm{h}})=0$.

The $k=1$ case corresponds to spherical symmetry, with the horizon having positive curvature, and this would be the only case permitted if the cosmological constant $\Lambda $ were either positive or zero. Considering instead $\Lambda<0$ opens up the possibility of black hole horizons of negative or vanishing curvature. 
For $k=1$, the coordinate $\theta $ is the usual spherical polar coordinate with values in the interval $[0,\pi)$. The metric in this case is the usual Schwarzschild anti-de Sitter spacetime.

When $k=0$, the event horizon has zero curvature and can be interpreted as a black ``sheet'' in the space-time.
In this case $(\theta, \varphi )$ are plane polar coordinates on the event horizon, $\theta $ corresponding to the radial distance from some fixed point on the horizon, so that $\theta \in [0,\infty )$.

In the final case, $k=-1$,  the horizon has constant negative curvature and can be thought of as a black hyperboloid.  
Here $\theta $ again takes on all positive real values.
When $k=-1$, there is a minimum event horizon radius, given by 
\begin{equation}
\label{eq:rcrit}
r_{\rm{h}}^{\rm {min}} = \frac{L}{\sqrt {3}}.
\end{equation}

When $k=0$ or $-1$, the event horizon is noncompact. 
By making appropriate identifications (see, for example, the discussion in \cite{Brill:1997mf}), it is possible to construct compact event horizons in this case.
For $k=0$, it is straightforward in Cartesian coordinates on the plane to make identifications to give either a cylindrical or toroidal event horizon topology.
For $k=-1$, the identifications cannot be realized in terms of coordinates in a straightforward way, but compact event horizons of genus greater than two can be constructed \cite{Brill:1997mf}. 
In this paper we consider only noncompact horizons when $k=0$ or $-1$.

For $k=0$, due to the vanishing curvature of the event horizon, there is no natural length scale associated with the horizon radius and the metric (\ref{eq:metric}) has two scaling symmetries. 
The first is the standard length rescaling
\begin{equation}
t\rightarrow \rho t, \qquad r\rightarrow \rho r, \qquad M \rightarrow \rho M, \qquad L \rightarrow \rho L
\label{eq:scaling1}
\end{equation}
where $\rho $ is an arbitrary constant.  
Under this rescaling $\rmd s^{2}\rightarrow \rho ^{2}\rmd s^{2}$ and $r_{\rm{h}}\rightarrow \rho r_{\rm{h}}$, with $\theta $ invariant.
The second rescaling has $L$ invariant and
\begin{equation}
t\rightarrow \rho ^{-1} t, \qquad 
r\rightarrow \rho r, \qquad \theta \rightarrow \rho ^{-1}\theta ,
\qquad M \rightarrow \rho ^{3}M .
\label{eq:scaling2}
\end{equation}
In this case $r_{\rm{h}}\rightarrow \rho r_{\rm{h}}$ but the metric $\rmd s^{2}$ is unchanged.
The two scaling symmetries (\ref{eq:scaling1}, \ref{eq:scaling2}) mean that, for $k=0$ it is sufficient to consider just one black hole metric (which could be taken to be, for example, the case $M=4$, $L=1$). 
Black holes with any other values of the parameters $M$ and $L$ and $k=0$ can be constructed from this particular black hole by appropriate application of the scalings  (\ref{eq:scaling1}, \ref{eq:scaling2}).

For fixed $k$, the black hole metric (\ref{eq:metric}) is parameterized by two quantities: the adS length scale $L$ and the black hole mass $M$.
The metric function $f(r)$ (\ref{eq:f}) has a single zero at $r=r_{\rm{h}}$, which depends on both $M$ and $L$ for each $k$.
To ease the comparison of our results for different black holes, it is helpful to change to a dimensionless radial coordinate for which the event horizon radius takes the same value irrespective of the parameters $M$ and $L$.
We start by defining a dimensionless parameter $\alpha $ which is related to $M$ and $L$ via
\begin{equation}
\alpha L\left( 
4\alpha ^{2}+k
\right) = M
\label{eq:alpha}
\end{equation}
and then define our new radial coordinate $\zeta $ by
\begin{equation}
\zeta = \frac{4\alpha ^{2}+k}{M}r - 1.
\label{eq:zeta}
\end{equation}
In terms of $\zeta $ the metric (\ref{eq:metric}) takes the form
\begin{equation}
\rmd s^{2} = -{\overline {f}}(\zeta ) \, \rmd t^{2} + \frac{\alpha ^{2}L^{2} \, \rmd \zeta ^{2}}{{\overline {f}}(\zeta )} + \alpha ^{2}L^{2} \left( \zeta + 1 \right) ^{2} \, \rmd \Omega _{k}^{2},
\end{equation}
with the metric function
\begin{equation}
{\overline {f}}(\zeta ) =f(r)= \frac{\zeta -1}{ \zeta + 1 } h(\zeta),\qquad h(\zeta)\equiv k+\alpha^{2}(\zeta^{2}+4\zeta+7).
\label{eq:fzeta}
\end{equation}
The event horizon $r=r_{\rm{h}}$ is then located at $\zeta =1$ for all $M$ and $L$, and the curvature singularity at $r=0$ is at $\zeta=-1$. For $k=0, 1$, the function $h(\zeta)$ has no real roots for all values of $\alpha$. Moreover, for $k=1$, the asymptotically flat Schwarzschild limit is given by $L\to\infty$ whence the parameter $\alpha$ necessarily vanishes and so $h(\zeta)=1$ in this case. For $k=0$, the scalings (\ref{eq:scaling1}, \ref{eq:scaling2}) do not change $\zeta $; under the first scaling $\alpha $ is invariant, whereas under the second scaling $\alpha \rightarrow \rho \alpha $. For $k=-1$, the function $h(\zeta)$ has no real roots as long as $\alpha^{2}>1/3$ and so we restrict to this range in this case. We will parameterize our black holes by $(k,\alpha, M)$ rather than $(k,L,M)$. As a final note in this section, we point out that in terms of this parametrization, the surface gravity of the black holes and the Ricci curvature scalar are, respectively,
\begin{eqnarray}
\label{eq:kappaR}
\kappa=\frac{1}{2}f'(r_{\mathrm{h}})=\frac{1}{4 M}(12 \alpha^{2}+k)(4 \alpha^{2}+k),
\nonumber\\
R=-\frac{12}{L^{2}}=-\frac{12 \alpha^{2}(4\alpha^{2}+k)^{2}}{M^{2}}.
\end{eqnarray}

\section{Quantum scalar fields on topological black hole space-times}
\label{sec:scalar}

We wish to consider the VP for a quantum Klein-Gordon field in the Hartle-Hawking state propagating on the (fixed) classical topological black hole space-times discussed in the previous section. We start with a classical field, $\phi(x)$, satisfying
\begin{equation}
\left\{\Box -\mu^{2}-\xi\,R\right\}\phi(x)=0,
\end{equation}
where $\Box$ is the D'Alembertian operator, $\mu$ is the field mass and $\xi$ is the coupling of the field to the background curvature. In the quantum theory, the field gets promoted to an operator-valued distribution, $\phi(x)\to\hat{\phi}(x)$, upon which is imposed a set of canonical commutation relations (see, for example, \cite{Birrell:1982ix}). The Feynman two-point function for a unit-norm state $|A\rangle$ is defined by
\begin{equation}
\label{eq:FeynmannDef}
-\rmi G_{{\mathrm {A}}}(x,x')=\langle {\mathrm {A}}| \mathcal{T}\{ \hat{\phi}(x)\hat{\phi} (x') \}|{\mathrm {A}}\rangle
\end{equation}
where $\mathcal{T}$ denotes the time-ordering operator, and $G_{{\mathrm {A}}}(x,x')$ is a Green's function for the Klein-Gordon operator, satisfying
\begin{equation}
\label{eq:KGEqn}
\left\{\Box -\mu^{2}-\xi\,R\right\}G_{{\mathrm {A}}}(x,x')=-\frac{\delta(x-x')}{\sqrt{|g|}}.
\end{equation}
Choosing a quantum state is tantamount to choosing a set of boundary conditions for this Green's function. Now the VP in a given quantum state $|{\mathrm {A}}\rangle$, which we denote by $\langle \hat{\phi}^{2}(x)\rangle_{{\mathrm {A}}}$, is provisionally defined by,
\begin{equation}
\label{eq:VPDef}
\langle \hat{\phi}^{2}(x)\rangle_{{\mathrm {A}}}=\rmi \lim_{x'\to x} G_{{\mathrm {A}}}(x,x').
\end{equation}
However, it is clear from (\ref{eq:KGEqn}--\ref{eq:VPDef}) that the VP is ill-defined since it involves the coincidence limit (that is, the limit $x'\to x$) of a solution to the wave equation whose source diverges in this limit. Extracting a meaningful finite quantity from this is the essence of the regularization problem which we discuss in the following section. For the remainder of this section, we will focus on constructing the Green's function in the Hartle-Hawking state $|{\mathrm {HH}}\rangle$.

Choosing to work in the Hartle-Hawking state is convenient since one can construct the Green's function in this case by working in the Euclidean sector. We perform a Wick rotation $t\to - \rmi \tau$ and enforce regular boundary conditions for the Green's function when one of the points is on the black hole horizon. Moreover, to avoid a conical singularity at the horizon, we must identify in the Euclidean time direction $\tau \rightarrow \tau+2\pi/\kappa$. This identification discretizes the frequency modes in a Fourier expansion of the Green's function which is one of the main advantages of working in the Euclidean sector. The (unregularized) VP for the scalar field in the Hartle-Hawking state is then
\begin{equation}
\langle \hat{\phi}^{2} \rangle_{\mathrm{HH}}^{\mathrm{unreg}}=\lim_{x'\to x}G_{\mathrm{E}}(x,x'),
\label{eq:VPunren}
\end{equation}
where $G_{\mathrm{E}}(x,x')$ denotes the Green's function on the Euclidean space-time subject to the appropriate regularity conditions.

Adopting standard separation of variable procedures for constructing the Green's function, we find that for all permissible $k$-values, the Green's function can be expressed as
\begin{equation}
\label{eq:EuclideanModeSum}
	G_{\mathrm{E}}(x,x')=\frac{\kappa}{(2\pi)^{2}}\sum_{n=-\infty}^{\infty}\rme ^{\rmi n \kappa \Delta \tau}\int_{\lambda=0}^{\infty}\rmd \lambda \,\mathcal{P}_{\lambda}^{(k)}(\gamma) g_{n\lambda}(r,r'),
\end{equation}
where
\begin{equation}
\mathcal{P}_{\lambda}^{(k)}(\gamma)=
\left\{ \begin{array}{@{\kern2.5pt}lL}
    \hfill (\lambda+\frac{1}{2})P_{\lambda}(\cos\gamma), & \qquad $k=1$,\\
    \hfill \lambda J_{0}(\lambda\,\gamma), & \qquad $k=0$,\\
           \lambda\,\tanh(\pi\lambda)\,P_{-\frac{1}{2}+\rmi\lambda}(\cosh\gamma), & \qquad $k=-1$.
\end{array}\right.
\label{eq:pdef}
\end{equation}
In (\ref{eq:pdef}), $P_{\lambda }(\cos \gamma )$ is a Legendre function, $J_{0}(\lambda, \gamma )$ a Bessel function of the first kind, $P_{-\frac{1}{2}+\rmi\lambda}(\cosh\gamma)$ is a conical (Mehler) function and
 $\gamma$ is the geodesic distance on $\rmd \Omega_{k}^{2}$ which is related to the coordinates $(\theta,\varphi)$ by
\begin{eqnarray}
	\label{eq:DefGamma}
	\cos\gamma= \cos\theta\,\cos\theta'+\sin\theta\,\sin\theta'\,\cos\Delta\varphi, &\quad k=1,\nonumber\\
	\gamma^{2} =\frac{1}{2}(\theta^{2}+\theta'^{2}-2 \theta\,\theta'\cos\Delta\varphi),  &\quad k=0,\nonumber\\
	\cosh\gamma= \cosh\theta\,\cosh\theta '-\sinh\theta\,\sinh\theta'\,\cos\Delta\varphi, &\quad k=-1,
\end{eqnarray}
where $\Delta \varphi = \varphi - \varphi '$.
In the spherical case, $k=1$, the eigenvalue $\lambda$ is an integer and the integral above should be replaced by a sum. Nevertheless, for simplicity of notation, we shall use $\int\,\rmd \lambda$ to denote all cases.
Many of the quantities considered in this paper depend on $k$; in general we do not explicitly include this dependence in our notation. 

The one-dimensional radial Green's function $g_{n\lambda}(r,r')$ satisfies the ODE
\begin{equation}
\fl	\Big\{\frac{\rmd }{\rmd r}\Big(r^{2}f(r)\frac{\rmd }{\rmd r}\Big)-\frac{n^{2}\kappa^{2}r^{2}}{f(r)}-\nu_{\lambda}-r^{2}(\mu^{2}+\xi\,R)\Big\}g_{n\lambda}(r,r')=-\delta(r-r'),
\label{eq:radialODE}
\end{equation}
where
\begin{equation}
	\nu_{\lambda}=\left[ \lambda+\sfrac{1}{4}k(k+1)\right] ^{2}-\sfrac{1}{4}k,\qquad k=0,\pm 1.
\end{equation}
In terms of the $\zeta$-coordinate, and using the explicit expressions (\ref{eq:kappaR}), we obtain
\begin{eqnarray}
&\Big\{\frac{\rmd }{\rmd \zeta}\Big[(\zeta^{2}-1)h(\zeta)\frac{\rmd }{\rmd \zeta}\Big]-\frac{n^{2}(12\alpha^{2}+k)^{2}(\zeta+1)^{4}}{16(\zeta^{2}-1)h(\zeta)}-\nu_{\lambda}\nonumber\\
&\quad-\alpha^{2}(\tilde{\mu}_{\xi}-2)(\zeta+1)^{2}\Big\}g_{n\lambda}(\zeta,\zeta')=-\frac{(4\alpha^{2}+k)}{M}\delta(\zeta-\zeta'),
\label{eq:ginhomo}
\end{eqnarray}
where we have defined the constant
\begin{equation}
\tilde{\mu}_{\xi}=\frac{M^{2}\mu^{2}}{\alpha^{2}(4\alpha^{2}+k)}-12(\xi-\sfrac{1}{6}).
\end{equation}
This definition implies that $\tilde{\mu}_{\xi}=0$ for massless, conformally-coupled scalar fields. Now, to solve the inhomogeneous equation (\ref{eq:ginhomo}), we take a normalized product of homogeneous solutions. In particular, let $p_{n\lambda}(\zeta)$ be an homogeneous solution that is regular on the horizon and $q_{n\lambda}(\zeta)$ be an homogeneous solution regular at infinity, then
\begin{equation}
	g_{n\lambda}(\zeta,\zeta')=\frac{(4\alpha^{2}+k)}{M}\frac{p_{n\lambda}(\zeta_{<})\,q_{n\lambda}(\zeta_{>})}{N_{n\lambda }}
    \label{eq:radialg}
\end{equation}
where $\zeta_{<}=\min\{\zeta,\zeta'\}$, $\zeta_{>}=\max\{\zeta,\zeta'\}$ and $N_{n\lambda }=-(\zeta^{2}-1)h(\zeta)W\{p_{n\lambda },q_{n\lambda }\}$ is the normalization constant constructed from the Wronskian of the two solutions. 

An important point in the construction of the mode-sum representation of the Green's function is that the pathological behaviour in the coincidence limit manifests as the nonconvergence of the mode-sums in (\ref{eq:EuclideanModeSum}). A meaningful way to apply an ultraviolet cut-off is addressed in the next section.

\section{Hadamard regularization}
\label{sec:hadamard}

The na\"{\i}ve expression for the VP (\ref{eq:VPunren}) is ill-defined and requires a prescription that (i) removes the singular terms (a process called regularization) and (ii) absorbs the terms introduced in order to cure the divergences into some other parameters in the theory (a process called renormalization). The conceptual framework for achieving this in a curved space-time, known as the point-splitting scheme, dates back to seminal work by DeWitt and Christensen \cite{DeWitt:1975ys,Christensen:1976vb,Christensen:1978yd}. A more general axiomatic variant of their original point-splitting prescription, known as Hadamard regularization \cite{Wald:1977up,Wald:1995yp,Decanini:2005eg}, is the approach adopted here.

\subsection{The Hadamard parametrix}
\label{sec:parametrix}

In general, the regularized VP for a scalar field in the Hartle-Hawking state is defined by
\begin{equation}
\label{eq:VPRegDef}
	\langle \hat{\phi}^{2} \rangle_{\mathrm{HH}}\equiv \lim_{x'\to x}\left[G_{\mathrm{E}}(x,x')-G_{\mathrm{S}}(x,x')\right],
\end{equation}
where $G_{\mathrm{S}}(x,x')$ is any symmetric two-point distribution which regularizes the Green's function in the coincidence limit, and which depends only on the geometry through the metric and its derivatives. The latter property guarantees that the regularization only introduces terms that can be reabsorbed elsewhere in the semi-classical field equations  (\ref{eq:scee}) (albeit with infinite renormalizations of higher-curvature terms). One particular family of two-point distributions with these properties are the Hadamard parametrices, 
\begin{equation}
\label{eq:GHadamard}
	G_{\mathrm{S}}(x,x')=\frac{1}{8\pi^{2}}\Big[\frac{\Delta^{1/2}(x,x')}{\sigma(x,x')}+V(x,x')\,\log(\sigma(x,x')/\ell^{2})\Big],
\end{equation}
where $\sigma(x,x')$ is Synge's world function, $\Delta^{1/2}(x,x')$ is the Van Vleck-Morette determinant, and $V(x,x')$ is a symmetric regular biscalar which is a homogeneous solution of the wave equation. For high-order covariant Taylor expansions for these biscalars, see  \cite{Decanini:2005eg}. The parameter $\ell$ is an arbitrary length-scale needed to make the argument of the $\log$ dimensionless. By construction, $G_{\mathrm{S}}(x,x')$ satisfies
\begin{equation}
\left\{\Box -\mu^{2}-\xi\,R\right\}G_{{\mathrm {S}}}(x,x')=-\frac{\delta(x-x')}{\sqrt{|g|}}+W(x,x'),
\end{equation}
where $W(x,x')$ is a geometric biscalar that is regular in the coincidence limit. This guarantees that the Green's function $G_{\mathrm{E}}(x,x')$ and the two-point distribution $G_{\mathrm{S}}(x,x')$ have the same short-distance singular structure, and hence the difference in (\ref{eq:VPRegDef}) is finite in the coincidence limit. 

This solves the conceptual problem of how to define the VP. However, computing the VP in practice remains a significant challenge. The crux of the difficulty is how to subtract the Hadamard parametrix (\ref{eq:GHadamard}) from the the mode-sum representation of the Green's function (\ref{eq:EuclideanModeSum}) in such a way that the limit can be meaningfully taken and the result can be numerically computed in a pragmatic way. A very efficient solution to this problem was recently devised by Taylor and Breen \cite{Taylor:2016edd,Taylor:2017sux} in static, spherically-symmetric space-times of arbitrary dimension. This involves expanding the Hadamard parametrix in a set of parameters adapted to the static, spherically-symmetric geometry and performing a simultaneous Fourier frequency and multipole decomposition of the terms in this expansion. The net result is that the Hadamard parametrix $G_{\mathrm{S}}(x,x')$ is expressed in the same set of basis mode functions as the Green's function and hence a mode-by-mode subtraction is amenable, resulting in a convergent mode-sum. Here we generalize this ``extended coordinates'' method to other horizon topologies. 

We proceed by simplifying matters by considering only a massless, conformally-coupled scalar field. In this case, in a short-distance expansion we have $V\sim\Or (\Delta x^{4})$, implying that the $\log$ term in (\ref{eq:GHadamard}) does not contribute in the coincidence limit. The task now reduces to obtaining a mode-sum representation of the direct part of the Hadamard parametrix $\Delta^{1/2}/\sigma$. Following \cite{Taylor:2016edd,Taylor:2017sux}, we start by defining the expansion parameters adapted to the geometry, which Taylor and Breen call ``extended coordinates''. For our purposes, the approriate extended coordinates are
\begin{eqnarray}
	\label{eq:ExpVar}
 w^{2} & = &\frac{2}{\kappa^{2}}(1-\cos \kappa\Delta\tau),
\nonumber \\
s^{2} &= &\left\{\begin{array}{ll}
    f(r)\,w^{2}+2 r^{2}(1-\cos\gamma), & k=1,\\
    f(r)\,w^{2}+2 r^{2}\gamma^{2}, & k=0,\\
    f(r)\,w^{2}+2 r^{2}(\cosh\gamma-1), & k=-1.
    \end{array}
    \right.
\end{eqnarray}
The extended coordinates $w$ and $s$ are formally treated as $\Or (\epsilon)\sim \Or (\Delta x)$ quantities. To leading order, the world function is given simply by $\sigma=\sfrac{1}{2}\epsilon^{2}(s^{2}+\Delta r^{2}/f)+\Or (\epsilon^{3})$, where here and henceforth we insert explicit powers of $\epsilon$ as a book-keeping mechanism for tracking the order of each term in the expansion. Higher order terms in the expansion are obtained by substituting the ansatz
\begin{equation}
	\label{eq:SigmaExp}
	\sigma=\sum_{ijk}\sigma_{ijk}(r)w^{i}\Delta r^{j}s^{k}\epsilon^{i+j+k}
\end{equation}
into the defining equation $\nabla_{a}\sigma \nabla^{a}\sigma=2\sigma$ and equating order by order to determine the coefficients $\sigma_{ijk}(r)$. A similar approach is used for the expansion of $\Delta^{1/2}(x,x')$. Since we are ultimately interested in the coincidence limit, we can make a further simplification by taking the partial coincidence limit $\Delta r=0$, then it can be shown that for all event horizon topologies being considered, 
the direct part of the Hadamard parametrix possesses an expansion of the form
\begin{eqnarray}
	\label{eq:HadamardDirectExp}
\frac{\Delta^{1/2}}{\sigma}&= &\sum_{i=0}^{m}\sum_{j=0}^{ i}\mathcal{D}^{(+)}_{ij}(r)\epsilon^{2i-2}\frac{w^{2i+2j}}{s^{2j+2}}
\nonumber \\ & & 
+\sum_{i=1}^{m}\sum_{j=1}^{ i}\mathcal{D}^{(-)}_{ij}(r)\epsilon^{2i-2}w^{2i-2j}s^{2j-2}+\Or (\epsilon^{2m}),
\end{eqnarray}
where $m$  is the truncation order, that is, we are ignoring terms that tend to zero in the coincidence limit at least as fast as $\epsilon^{2m}$. This is not true for the Hadamard parametrix $G_{\mathrm{S}}(x,x')$ as a whole since the terms we are ignoring in the tail term $V(x,x')\,\log[\sigma (x,x')/\ell ^{2}]$ tend to zero in the coincidence limit like $\Or (\epsilon^{4}\,\log\epsilon)$ for a massless, conformally-coupled scalar field. Hence, taking the truncation order to be $m=2$ above, we have
\begin{eqnarray}
	\label{eq:HadamardExp}
\fl	G_{\mathrm{S}}(\Delta\tau,\gamma,r)=\frac{1}{8\pi^{2}}\left(\sum_{i=0}^{2}\sum_{j=0}^{ i}\mathcal{D}^{(+)}_{ij}(r)\epsilon^{2i-2}\frac{w^{2i+2j}}{s^{2j+2}}+\sum_{i=1}^{2}\sum_{j=1}^{ i}\mathcal{D}^{(-)}_{ij}(r)\epsilon^{2i-2}w^{2i-2j}s^{2j-2}\right)\nonumber\\
+\Or (\epsilon^{4}\log\epsilon).
\end{eqnarray}
The goal is to decompose this two-point function in the same basis modes as the corresponding Euclidean Green's function for each horizon topology. This allows us to define the difference in (\ref{eq:VPRegDef}) mode by mode. The resultant mode-sum would converge even if we truncated the expansion of $G_{\mathrm{S}}(x,x')$ in (\ref{eq:HadamardExp}) at $i=0$, though the convergence would be extremely slow and only conditional. Including the $i=1$ and $i=2$ terms serves to speed up the convergence by capturing more of the high-frequency and high-multipole behaviour of the Green's function. However, the terms involving $\mathcal{D}_{ij}^{(-)}(r)$ involve only even positive powers of $w$ and $s$, and hence are polynomial in $\cos\kappa\Delta\tau$. Decomposing such terms in Fourier frequency modes would result in a sum over finite frequencies. That is, these terms do not have a large-frequency contribution and cannot affect the convergence. Similarly, these terms do not have a large-multipole $\lambda$ contribution. This implies that it is redundant to decompose terms involving $\mathcal{D}_{ij}^{(-)}(r)$ in the Hadamard parametrix since they cannot improve the convergence of the resultant mode-sum, because they do not contribute for large $\lambda$ and $n$. These terms are best kept in closed form. Moreover, since we are eventually interested in the coincidence limit, only the zeroth order polynomial survives, that is, the $\mathcal{D}_{ij}^{(-)}$ term with $i=j=1$. Given $\mathcal{D}_{11}^{(-)}(r)=-f'(r)/(6r)$, we may re-express (\ref{eq:HadamardExp}) as
\begin{eqnarray}
	\label{eq:HadamardExpFinal}
	G_{\mathrm{S}}(\Delta\tau,\gamma,r)=\frac{1}{8\pi^{2}}\left(\sum_{i=0}^{2}\sum_{j=0}^{ i}\mathcal{D}^{(+)}_{ij}(r)\epsilon^{2i-2}\frac{w^{2i+2j}}{s^{2j+2}}-\frac{f'(r)}{6r}\right)+\cdots
\end{eqnarray}
The dots in this expression now represent the $\Or (\epsilon^{4}\log\epsilon)$ terms that we are ignoring but also some terms that are formally of lower order but which neither contribute in the coincidence limit nor improve the convergence of the mode-sum in the VP.
The coefficients ${\mathcal {D}}^{(+)}_{ij}$ are given in Table \ref{tab:Dcoeff}.
\begin{table}[h]
\centering
 	\begin{tabular}{|c|c|} \hline
 		& $\mathcal{D}^{(+)}_{ij}(r)$ coefficients for topological black holes\\ \hline \hline & \\
 $\mathcal{D}_{00}^{(+)}(r)$& $2$  \\ &\textit{} \\\hline & \\
 $\mathcal{D}_{10}^{(+)}(r)$& $\displaystyle{-\frac{f(r)\left[-2k+2 f(r)-2 r f'(r)+r^{2}f''(r)\right]}{12 r^{2}}}$  \\ &\\
 $\mathcal{D}_{11}^{(+)}(r)$& $\displaystyle{\frac{f(r)\left[4 f(r)^{2}+f(r)(-4 k -4 r f'(r))+r^{2}(-4\kappa^{2}+f'(r)^{2})\right]}{24 r^{2}}}$ 
      \\ &\\\hline &\\
 $\mathcal{D}_{20}^{(+)}(r)$& $\begin{array}{c}\displaystyle{\frac{1}{2880
   r^{4}}}f(r)\Big\{76 f(r)^{3}-5 r^{2}\left[4\kappa^{2}-f'(r)^{2}\right]\left[ -2k-2 r f'(r)+r^{2}f''(r)\right] \\
   -8 f(r)^{2}\left[10 k +19 r f'(r)-7 r^{2}f''(r)+3 r^{3}f'''(r)\right] \\
   +f(r)\left[4-40 r^{2}\kappa^{2}+86 r^{2}f'(r)^{2}-20 k r^{2}f''(r)+9 r^{4}f''(r)^{2} \right. \\ \left.
   +4 r f'(r)(20 k -14 r^{2}f''(r)+3 r^{3}f'''(r))\right]\Big\}\end{array}$ 
     \\ & \\
 $\mathcal{D}_{21}^{(+)}(r)$& $\begin{array}{c}\displaystyle{-\frac{1}{2880
   r^{4}}}f(r)\Big\{104 f(r)^{4}+r^{4}\left[ 64\kappa^{2}-20\kappa^{2}f'(r)^{2}+f'(r)^{4}\right] \\
   +4 f(r)^{3}\left[ -40 k -52 r f'(r)+11 r^{2}f''(r)\right]+r^{2}f(r)\left[ 120 r \kappa^{2}f'(r) \right. \\ \left.
   -30 r f'(r)^{3}-20\kappa^{2}(-6 k +r^{2}f''(r))+f'(r)^{2}(-30 k+11 r^{2}f''(r)) \right]\\-2 f(r)^{2}\left[ -28+60 r^{2}\kappa^{2}-67 r^{2}f'(r)^{2}+10 k r^{2}f''(r) \right. \\ \left. +f'(r)(-80 k r+22 r^{3}f''(r))\right] \Big\}\end{array}$ 
     \\ & \\
$\mathcal{D}_{22}^{(+)}(r)$& $\begin{array}{c}\displaystyle{\frac{f(r)^{2}\left[ 4 f(r)^{2}+f(r)(-4k-4 r f'(r))+r^{2}(-4\kappa^{2}+f'(r)^{2})\right] ^{2}}{1152 r^{4}}}\end{array}$ 
     \\ &  \\\hline
    \end{tabular}
    \caption{Hadamard coefficients $\mathcal{D}_{ij}^{(+)}(r)$ for the topological Schwarzschild-adS black hole space-times. The parameter $k$ assumes the values $k=1,0,-1$ corresponding to spherical, planar and hyperboloidal horizons, respectively.}
    \label{tab:Dcoeff}
 \end{table}

To express $G_{\mathrm{S}}(x,x')$ as a mode-sum, we assume an ansatz of the form 
\begin{equation}
\label{eq:RegParamAnsatz}
\frac{w^{2i+2j}}{s^{2j+2}}=\sum_{n=-\infty}^{\infty}\rme ^{\rmi n \kappa\Delta\tau}\int_{\lambda=0}^{\infty}\mathcal{P}_{\lambda}^{(k)}(\gamma)\Psi_{n\lambda}(i,j|r) \, \rmd \lambda
\end{equation}
which is similar to the mode-sum expression for the Euclidean Green's function $G_{\rm {E}}(x,x')$ (\ref{eq:EuclideanModeSum}).
The $\Psi_{n\lambda}(i,j|r)$ are known as regularization parameters. To compute the regularization parameters, we must invert (\ref{eq:RegParamAnsatz}), which is achieved by multiplying across by $\mathcal{P}_{\lambda'}^{(k)}(\gamma) \rme ^{-\rmi n'\kappa\Delta\tau}$ and integrating, applying the orthogonality relations
\begin{equation}
\int_{0}^{2\pi/\kappa}\rme ^{\rmi (n-n')\kappa \Delta\tau}\rmd \Delta\tau=\frac{2\pi}{\kappa}\delta_{n n'},
\end{equation}
and
\begin{equation}
\int_{0}^{(\pi,\infty)}\mathcal{P}^{(k)}_{\lambda}(\gamma)\mathcal{P}^{(k)}_{\lambda'}(\gamma)\mathcal{F}_{k}(\gamma)\,\rmd \gamma=\beta_{k}(\lambda)\delta(\lambda-\lambda'),
\label{eq:calPintegral}
\end{equation}
where ${\mathcal {F}}_{k}(\gamma )$ is given in (\ref{eq:Fk}).
In the integral (\ref{eq:calPintegral}), the upper bound is infinity for the non-compact cases $k=0,-1$ and $\pi$ for the compact case $k=1$. Also, the eigenvalue $\lambda$ is discrete for $k=1$ and the Dirac delta should be replaced by a Kronecker delta. The factor $\beta_{k}(\lambda)$ is
\begin{equation}
\beta_{k}(\lambda)=\left\{\begin{array}{ll}
\lambda+\sfrac{1}{2}, & k=1,\\
\lambda, & k=0,\\
\lambda\,\tanh(\pi\lambda), & k=-1.
\end{array}\right.
\end{equation}
The result of inverting (\ref{eq:RegParamAnsatz}) is the following double integral representation for the regularization parameters
\begin{equation}
\label{eq:RegParamIntegral}
\fl\Psi_{n\lambda}(i,j|r)=\frac{\kappa}{2\pi\beta_{k}(\lambda)}\int_{0}^{2\pi/\kappa}\int_{0}^{(\pi,\infty)}\rme ^{-\rmi  n\kappa\Delta\tau}\mathcal{P}_{\lambda}^{(k)}(\gamma)\frac{w^{2i+2j}}{s^{2j+2}}\mathcal{F}_{k}(\gamma) \,\rmd \gamma\, \rmd \Delta\tau.
\end{equation}
We next derive explicit representations for these double integrals in terms of known functions by considering each event horizon topology separately.

\subsection{$k=1$ regularization parameters}

For spherical horizons, the integral over $\gamma$ in (\ref{eq:RegParamIntegral}) assumes the form
\begin{equation}
\fl \int_{0}^{\pi}\mathcal{P}_{\lambda}^{(1)}(\gamma)\frac{w^{2i+2j}}{s^{2j+2}}\mathcal{F}_{1}(\gamma) \, \rmd \gamma=\frac{(\lambda+\frac{1}{2})2^{i-1}}{\kappa^{2i+2j}r^{2j+2}}(1-\cos\kappa\Delta\tau)^{i+j}\int_{0}^{\pi}\frac{P_{\lambda}(\cos\gamma)\,\sin\gamma\, \rmd \gamma}{(z-\cos\gamma)^{j+1}},
 \end{equation}
 where
 \begin{equation}
 z=1+\frac{f(r)}{2 r^{2}}w^{2}.
 \end{equation}
 Further defining
 \begin{equation}
 \eta=\sqrt{1+\frac{f(r)}{\kappa^{2}r^{2}}},
 \end{equation}
 and noting that
 \begin{equation}
 \left(\frac{\partial}{\partial z}\right)^{j}=\frac{1}{2^{j}(1-\cos\kappa\Delta\tau)^{j}}\left(\frac{1}{\eta}\frac{\partial}{\partial\eta}\right)^{j},
 \end{equation}
we obtain
 \begin{eqnarray}
 \label{eq:GammaIntSpherical}
& & \hspace{-2cm} \int_{0}^{\pi}\mathcal{P}_{\lambda}^{(1)}(\gamma)\frac{w^{2i+2j}}{s^{2j+2}}\mathcal{F}_{1}(\gamma)\, \rmd \gamma  
\nonumber \\
&  & =   \frac{(\lambda+\frac{1}{2})2^{i-j-1}(-1)^j}{\kappa^{2i+2j}r^{2j+2}j!}(1-\cos\kappa\Delta\tau)^{i}\left(\frac{1}{\eta}\frac{\partial}{\partial\eta}\right)^{j}\int_{0}^{\pi}\frac{P_{\lambda}(\cos\gamma)\sin\gamma\,\rmd \gamma}{(z-\cos\gamma)}\nonumber\\
& & = \frac{(\lambda+\frac{1}{2})2^{i-j}(-1)^j}{\kappa^{2i+2j}r^{2j+2}j!}(1-\cos\kappa\Delta\tau)^{i}\left(\frac{1}{\eta}\frac{\partial}{\partial\eta}\right)^{j}Q_{\lambda}(z),
 \end{eqnarray}
 where the last line follows from a standard integral identity for the Legendre functions. This must be substituted into (\ref{eq:RegParamIntegral}) and the time integral must be performed, keeping in mind that there is a time-dependence in the definition of $z$ that appears in the argument of the Legendre function of the second kind. In order to do the time integral analytically, we use the Legendre addition theorem,
 \begin{equation}
 \label{eq:AdditionSpherical}
 Q_{\lambda}(z)=\sum_{p=-\infty}^{\infty}(-1)^{p}\rme ^{\rmi p \kappa\Delta\tau}P_{\lambda}^{-|p|}(\eta)Q_{\lambda}^{|p|}(\eta).
 \end{equation}
 Combining equations (\ref{eq:GammaIntSpherical}) and (\ref{eq:AdditionSpherical}) in equation (\ref{eq:RegParamIntegral}) gives
 \begin{eqnarray}
 \Psi_{n\lambda}(i,j|r)=\frac{\kappa}{2\pi}&\frac{2^{i-j}(-1)^{j}}{\kappa^{2i+2j}r^{2j+2}j!}\sum_{p=-\infty}^{\infty}(-1)^{p}\left(\frac{1}{\eta}\frac{\partial}{\partial \eta}\right)^{j}P_{\lambda}^{-|p|}(\eta)Q_{\lambda}^{|p|}(\eta)\nonumber\\
& \times \int_{0}^{2\pi/\kappa} \rme ^{\rmi (p-n)\kappa\Delta\tau}(1-\cos\kappa\Delta\tau)^{i} \, \rmd \Delta\tau.
 \end{eqnarray}
 The time integral is now straightforward,
 \begin{equation}
 \label{eq:TimeIntegral}
 \int_{0}^{2\pi/\kappa} \rme ^{\rmi (p-n)\kappa\Delta\tau}(1-\cos\kappa\Delta\tau)^{i} \, \rmd \Delta\tau=\frac{2\pi}{\kappa}\frac{i!(2i-1)!!(-1)^{p-n}}{(i-n+p)!(i+n-p)!} ,
 \end{equation}
 whence the regularization parameters reduce to
 \begin{equation}
\fl \Psi_{n\lambda}(i,j|r)=\frac{2^{i-j}i!(2i-1)!!(-1)^{n+j}}{\kappa^{2i+2j}r^{2j+2}j!}\sum_{p=n-i}^{n+i}\left(\frac{1}{\eta}\frac{\partial}{\partial \eta}\right)^{j}\frac{P_{\lambda}^{-|p|}(\eta)\,Q_{\lambda}^{|p|}(\eta)}{(i-n+p)!(i+n-p)!}.
 \end{equation}
The truncation of the infinite sum is a result of the presence of the factorials in the denominator of (\ref{eq:TimeIntegral}).

\subsection{$k=0$ regularization parameters}
For planar horizons, the integral over $\gamma$ in (\ref{eq:RegParamIntegral}) assumes the form
\begin{equation}
\fl \int_{0}^{\infty}\mathcal{P}_{\lambda}^{(0)}(\gamma)\frac{w^{2i+2j}}{s^{2j+2}}\mathcal{F}_{0}(\gamma) \, \rmd \gamma=\frac{\lambda\,2^{i-1}}{\kappa^{2i+2j}r^{2j+2}}(1-\cos\kappa\Delta\tau)^{i+j}\int_{0}^{\infty}\frac{J_{0}(\lambda\gamma)\,\gamma\,\rmd \gamma}{(z+\gamma^{2})^{j+1}},
 \end{equation}
 where now we have
 \begin{equation}
 z=\frac{f(r)}{2 r^{2}}w^{2}.
 \end{equation}
Defining
 \begin{equation}
 \eta=\sqrt{\frac{f(r)}{\kappa^{2}r^{2}}},
 \end{equation}
 implies that
 \begin{equation}
 \left(\frac{\partial}{\partial z}\right)^{j}=\frac{1}{2^{j}(1-\cos\kappa\Delta\tau)^{j}}\left(\frac{1}{\eta}\frac{\partial}{\partial\eta}\right)^{j}
 \end{equation}
and hence we obtain
 \begin{eqnarray}
 \label{eq:GammaIntPlanar}
\fl\int_{0}^{\infty}\mathcal{P}_{\lambda}^{(0)}(\gamma)\frac{w^{2i+2j}}{s^{2j+2}}\mathcal{F}_{0}(\gamma) \, \rmd \gamma&=\frac{\lambda\,2^{i-j-1}(-1)^{j}}{\kappa^{2i+2j}r^{2j+2}j!}(1-\cos\kappa\Delta\tau)^{i}\left(\frac{1}{\eta}\frac{\partial}{\partial\eta}\right)^{j}\int_{0}^{\infty}\frac{J_{0}(\lambda\gamma) \, \gamma\,\rmd \gamma}{(z+\gamma^{2})}\nonumber\\
&=\frac{\lambda\,2^{i-j-1}(-1)^{j}}{\kappa^{2i+2j}r^{2j+2}j!}(1-\cos\kappa\Delta\tau)^{i}\left(\frac{1}{\eta}\frac{\partial}{\partial\eta}\right)^{j}K_{0}(\lambda\sqrt{z}),
 \end{eqnarray}
 where the last line follows from a standard integral identity for the modified Bessel functions. As before, we factor out the time dependence using an appropriate addition theorem, in this case, we employ the identity
 \begin{equation}
 \label{eq:AdditionPlanar}
 K_{0}(\lambda\sqrt{z})=\sum_{p=-\infty}^{\infty}\rme ^{\rmi p \kappa\Delta\tau}I_{p}(\lambda\,\eta)K_{p}(\lambda\,\eta).
 \end{equation}
 Combining equations  (\ref{eq:GammaIntPlanar}) and (\ref{eq:AdditionPlanar}) in equation (\ref{eq:RegParamIntegral}) gives
 \begin{eqnarray}
 \Psi_{n\lambda}(i,j|r)=\frac{\kappa}{2\pi}&\frac{2^{i-j-1}(-1)^{j}}{\kappa^{2i+2j}r^{2j+2}j!}\sum_{p=-\infty}^{\infty}\left(\frac{1}{\eta}\frac{\partial}{\partial \eta}\right)^{j}I_{p}(\lambda\,\eta)K_{p}(\lambda\,\eta)\nonumber\\
& \times \int_{0}^{2\pi/\kappa} \rme ^{\rmi (n-p)\kappa\Delta\tau}(1-\cos\kappa\Delta\tau)^{i} \, \rmd \Delta\tau.
 \end{eqnarray}
 The time integral here is precisely the same as in the spherical case (\ref{eq:TimeIntegral}) and so the result for the regularization parameters is
 \begin{equation}
 \fl\Psi_{n\lambda}(i,j|r)=\frac{2^{i-j-1}i!(2i-1)!!(-1)^{n}}{\kappa^{2i+2j}r^{2j+2}j!}\sum_{p=n-i}^{n+i}\left(\frac{1}{\eta}\frac{\partial}{\partial \eta}\right)^{j}\frac{(-1)^{p+j}I_{p}(\lambda\,\eta)K_{p}(\lambda\,\eta)}{(i-n+p)!(i+n-p)!}.
 \end{equation}

\subsection{$k=-1$ regularization parameters}
For hyperboloidal horizons, the integral over $\gamma$ in (\ref{eq:RegParamIntegral}) is
\begin{eqnarray}
\fl \int_{0}^{\infty}\mathcal{P}_{\lambda}^{(-1)}(\gamma)\frac{w^{2i+2j}}{s^{2j+2}}\mathcal{F}_{-1}(\gamma) \, \rmd \gamma=&\frac{\lambda\,\tanh(\pi \lambda) \, 2^{i-1}}{\kappa^{2i+2j}r^{2j+2}}(1-\cos\kappa\Delta\tau)^{i+j}\nonumber\\
&\times \,\int_{0}^{\infty}\frac{P_{-\frac{1}{2}+\rmi\lambda}(\cosh\gamma)\,\sinh\gamma\,\rmd \gamma}{(z+\cosh\gamma)^{j+1}},
 \end{eqnarray}
 where now we have
 \begin{equation}
 z=-1+\frac{f(r)}{2 r^{2}}w^{2}.
 \end{equation}
In this case, we define
 \begin{equation}
 \eta=\sqrt{1-\frac{f(r)}{\kappa^{2}r^{2}}},
 \end{equation}
 so that
 \begin{equation}
 \left(\frac{\partial}{\partial z}\right)^{j}=\frac{(-1)^{j}}{2^{j}(1-\cos\kappa\Delta\tau)^{j}}\left(\frac{1}{\eta}\frac{\partial}{\partial\eta}\right)^{j}.
 \end{equation}
Hence, we obtain
 \begin{eqnarray}
 \label{eq:GammaIntHyperboloid}
\fl\int_{0}^{\infty}\mathcal{P}_{\lambda}^{(-1)}(\gamma)\frac{w^{2i+2j}}{s^{2j+2}}\mathcal{F}_{-1}(\gamma) \, \rmd \gamma &
\nonumber \\
& \hspace{-4cm} 
=\frac{\lambda\,\tanh(\pi\lambda) \, 2^{i-j-1}}{\kappa^{2i+2j}r^{2j+2}j!}(1-\cos\kappa\Delta\tau)^{i}\left(\frac{1}{\eta}\frac{\partial}{\partial\eta}\right)^{j}
\int_{0}^{\infty}\frac{P_{-\frac{1}{2}+\rmi\lambda}(\cosh\gamma)\sinh\gamma\,\rmd \gamma}{(z+\cosh\gamma)}\nonumber\\
& \hspace{-4cm}
=\frac{\lambda\,\pi\,\tanh(\pi\lambda) \, 2^{i-j-1}}{\cosh(\pi\lambda) \, \kappa^{2i+2j}r^{2j+2}j!}(1-\cos\kappa\Delta\tau)^{i}\left(\frac{1}{\eta}\frac{\partial}{\partial\eta}\right)^{j}P_{-\frac{1}{2}+\rmi\lambda}(z),\nonumber\\
 \end{eqnarray}
where the last line follows from a standard integral identity for the conical (Mehler) functions. Factoring out the time dependence using an appropriate addition theorem is a little more subtle than in the other two cases. The first subtlety is associated with the range of the variable $\eta$, which in this case is related to $z$ explicitly by $z=-\eta^{2}-(1-\eta^{2})\cos\kappa\Delta\tau$. Recall that for $k=-1$, we have restricted to the range $\alpha^{2}> 1/3$,  which is sufficient to guarantee that $0<\eta^{2}\le 1$. To see this, we note that $\eta^{2}$ is a decreasing function of $r$ and, using the asymptotic values at the horizon and infinity, satisfies the inequality
\begin{equation}
1-\frac{1}{\kappa^{2}L^{2}}\le\eta^{2}\le 1.
\end{equation}
Requiring that $1-1/(\kappa^{2}L^{2})\ge 0$ is equivalent to $(12\alpha^{2}-1)^{2}\ge 16\alpha^{2}$ using (\ref{eq:kappaR}) and (\ref{eq:alpha}). Solving the latter inequality, it can be shown that it is certainly satisfied whenever $\alpha^{2}> 1/3$. With this range in mind, we can employ the addition theorem
 \begin{eqnarray}
 \label{eq:AdditionHyperboloid}
 & & \hspace{-2cm} P_{-\frac{1}{2}+\rmi\lambda}(-\cos\psi\,\cos\psi'-\sin\psi\,\sin\psi'\cos\kappa\Delta\tau)
 \nonumber \\ & & 
 =\sum_{p=-\infty}^{\infty}\rme ^{\rmi p \kappa\Delta\tau}P^{-p}_{-\frac{1}{2}+\rmi\lambda}(-\cos\psi)P^{p}_{-\frac{1}{2}+\rmi\lambda}(\cos\psi') 
 \end{eqnarray}
 with $\eta=\cos\psi=\cos\psi'$. The second subtlety in the hyperboloidal case is that this addition theorem is strictly only valid when $0<\psi'<\psi<\pi$ in the sense that the sum does not converge in the limit $\psi'\to\psi$ (or equivalently $\eta'\to\eta$). This is unlike the other cases where the appropriate addition theorems involved convergent sums, albeit sums that are only slowly converging. However, the convergence properties of the sum in (\ref{eq:AdditionHyperboloid}) turn out to be irrelevant since the time integral sets to zero all the terms in the infinite sum apart from a finite number in the range $p\in\{n-i,...,n+i\}$. A more rigorous approach would be to consider (\ref{eq:AdditionHyperboloid}) with $\psi'<\psi$, then perform the time integral before taking the limit $\psi'\to\psi$. The result is however the same as that obtained by using (\ref{eq:AdditionHyperboloid}) with $\psi=\psi'$ and ignoring the fact that the sum does not converge. Finally, by combining equations  (\ref{eq:GammaIntHyperboloid}) and (\ref{eq:AdditionHyperboloid})  in equation (\ref{eq:RegParamIntegral}) and performing the time integral using (\ref{eq:TimeIntegral}) gives
 \begin{eqnarray}
 \fl\Psi_{n\lambda}(i,j|r)=\frac{\pi\,2^{i-j-1}i!(2i-1)!!(-1)^{n}}{\cosh(\pi\lambda)\, \kappa^{2i+2j}r^{2j+2}j!}\sum_{p=n-i}^{n+i}\left(\frac{1}{\eta}\frac{\partial}{\partial \eta}\right)^{j}\frac{(-1)^{p}P_{-\frac{1}{2}+\rmi\lambda}^{-p}(-\eta)P^{p}_{-\frac{1}{2}+\rmi\lambda}(\eta)}{(i-n+p)!(i+n-p)!}.\nonumber\\
 \end{eqnarray}

\subsection{Mode-sum representation of the Hadamard parametrix}
\label{sec:hadsummary}

To summarize the results of this section succinctly, we have shown that for each event horizon topology considered, the Hadamard parametrix has a mode-sum representation of the form
\begin{eqnarray}
\fl G_{\mathrm{S}}(\Delta\tau, \gamma,r)=\frac{1}{8\pi^{2}}\sum_{n=-\infty}^{\infty}\rme ^{\rmi n \kappa\Delta\tau}\int_{0}^{\infty} \rmd \lambda \, \mathcal{P}^{(k)}_{\lambda}(\gamma)\sum_{i=0}^{2}\sum_{j=0}^{i}\mathcal{D}_{ij}^{(+)}(r)\Psi_{n\lambda}(i,j|r)-\frac{f'(r)}{48\pi^{2}r}\nonumber\\
\label{eq:modeHad}
\end{eqnarray}
where the coefficients $\mathcal{D}_{ij}^{(+)}(r)$ are given in Table \ref{tab:Dcoeff} and the regularization parameters $\Psi_{n\lambda}(i,j|r)$ are
\begin{eqnarray}
\fl\Psi_{n\lambda}(i,j|r)=\frac{2^{i-j}i!(2i-1)!!(-1)^{n}}{\kappa^{2i+2j}r^{2j+2}j!}\sum_{p=n-i}^{n+i}\left(\frac{1}{\eta}\frac{\partial}{\partial \eta}\right)^{j}\frac{\chi_{p\lambda}(\eta)}{(i-n+p)!(i+n-p)!}
\label{eq:Psifinal}
\end{eqnarray}
with
\begin{equation}
\chi_{p\lambda}(\eta)=\left\{\begin{array}{l l}
(-1)^{j}P^{-|p|}_{\lambda}(\eta)Q^{|p|}_{\lambda}(\eta), & k=1,\\ \\
\frac{1}{2}(-1)^{p+j}I_{p}(\lambda\eta)K_{p}(\lambda\eta), & k=0,\\ \\
{\displaystyle {\frac{\pi}{2\cosh(\pi\lambda)}}}(-1)^{p}P_{-\frac{1}{2}+\rmi\lambda}^{-p}(-\eta)P_{-\frac{1}{2}+\rmi\lambda}^{p}(\eta), & k=-1,\end{array}\right.
\end{equation}
and $\eta$ is defined by
\begin{equation}
\eta=\sqrt{\left|k+\frac{f(r)}{\kappa^{2}r^{2}}\right|}.
\end{equation}

\section{Renormalized vacuum polarization}
\label{sec:renVP}

In the previous section we derived a mode-sum representation of the Hadamard parametrix (\ref{eq:modeHad}), in terms of which the renormalized VP (\ref{eq:VPRegDef}) is
\begin{eqnarray}	
\hspace{-0.5cm}\langle \hat{\phi}^{2} \rangle_{\mathrm{HH}} & \equiv & \lim_{x'\to x}\left[G_{\mathrm{E}}(x,x')-G_{\mathrm{S}}(x,x')\right]
    \nonumber \\ 
  & = & 
\lim_{x'\to x} \left\{  \frac{1}{4\pi ^{2}}\sum_{n=-\infty}^{\infty}\rme ^{\rmi n \kappa \Delta \tau}\int_{\lambda=0}^{\infty}\rmd \lambda \,\mathcal{P}_{\lambda}^{(k)}(\gamma)
\right. \nonumber \\ & & \left. \times 
\left[ \kappa g_{n\lambda}(r,r')
-\frac{1}{2}\sum_{i=0}^{2}\sum_{j=0}^{i}\mathcal{D}_{ij}^{(+)}(r)\Psi_{n\lambda}(i,j|r)
\right] -\frac{f'(r)}{48\pi^{2}r} 
\right\}  
\nonumber \\ & = & 
\lim_{\gamma \to 0} \left\{  \frac{1}{4\pi ^{2}}\int_{\lambda=0}^{\infty}\rmd \lambda 
\sum_{n=-\infty}^{\infty}
\,\mathcal{P}_{\lambda}^{(k)}(\gamma)
\right. \nonumber \\ & & \left. \times 
\left[ \kappa g_{n\lambda}(\zeta )
-\frac{1}{2}\sum_{i=0}^{2}\sum_{j=0}^{i}\mathcal{D}_{ij}^{(+)}(r)\Psi_{n\lambda}(i,j|r)
\right] -\frac{f'(r)}{48\pi^{2}r} 
\right\}  
\nonumber \\ & = & 
\frac{1}{4\pi ^{2}}\int_{\lambda=0}^{\infty}\rmd \lambda \sum_{n=-\infty}^{\infty}\,\mathcal{P}_{\lambda}^{(k)}(0)
\left[ \kappa g_{n\lambda}(\zeta )
-\frac{1}{2}\sum_{i=0}^{2}\sum_{j=0}^{i}\mathcal{D}_{ij}^{(+)}(r)\Psi_{n\lambda}(i,j|r)
\right]
\nonumber \\ & & -\frac{f'(r)}{48\pi^{2}r}, 
\label{eq:VPfinal}
\end{eqnarray}
where we have defined $g_{n\lambda }(\zeta )= \lim _{r'\rightarrow r}g_{n\lambda }(r,r')$ and $g_{n\lambda }(r,r')$ is given by  (\ref{eq:radialg}).
To compute the renormalized VP numerically, we therefore need to find the mode functions $p_{n\lambda }(\zeta )$, $q_{n\lambda }(\zeta )$, the normalization constants $N_{n\lambda }$, and the regularization parameters $\Psi _{n\lambda }(i,j|r)$, before combining all these quantities into the sums in (\ref{eq:VPfinal}). 
We describe our numerical methodology for each of these parts before discussing our results. 

\subsection{Radial functions}
\label{sec:radial}

The radial functions $p_{n\lambda }(\zeta )$ and $q_{n\lambda }(\zeta )$ satisfy the homogeneous version of the ODE (\ref{eq:radialODE})
subject to the boundary conditions that $p_{n\lambda }$ is regular on the event horizon $\zeta =1$ and $q_{n\lambda }$ is regular as $\zeta \rightarrow \infty $.
The radial equation (\ref{eq:radialODE}) is invariant under the mapping $n\rightarrow -n$, so we only need to find the radial functions for $n\ge 0$.

Close to the horizon (which is a regular singular point of the ODE), we assume a Frobenius series expansion for $p_{n\lambda }$ of the form
\begin{equation}
p_{n\lambda  }(\zeta ) = \sum _{i=0}^{\infty } a_{i}\left(  \zeta - 1\right) ^{(c_{p}+i)},
\label{eq:pseries}
\end{equation}
where we set $a_{0}$ to be a convenient nonzero constant.
The indicial equation for $c_{p}$ has roots $c_{p}= \pm n/2$, therefore we take $c_{p}= \left|n \right| /2$, so that $p_{n\lambda }$ is regular on the horizon.
Since the ODE (\ref{eq:radialODE}) is singular at the horizon, we start our integration close to the horizon and use the Frobenius series (\ref{eq:pseries}) to give the initial values of $p_{n\lambda }$ and its first derivative.
We then integrate outwards to large $\zeta $.

Unlike the asymptotically flat case, infinity is a regular singular point of the ODE (\ref{eq:radialODE}). Thus for large $\zeta$, we can assume a Frobenius series for $q_{n\lambda}(\zeta)$ of the form
\begin{equation}
q_{n\lambda  }(\zeta ) = \sum _{i=0}^{\infty }b_{i}\zeta ^{-c_{q}-i},
\label{eq:qseries}
\end{equation}
where we set $b_{0}$ to be a convenient nonzero constant.
The indicial equation for $c _{q}$ has roots $c_{q}=1$, $2$ when the scalar field is massless and conformally coupled, which means that either choice will give $q_{n\lambda }$ regular as $\zeta \rightarrow \infty $.
Since adS space-time is not globally hyperbolic, it is necessary to impose boundary conditions on a quantum scalar field at time-like infinity in order for the time-evolution of the field to be well-defined \cite{Avis:1977yn}.
On pure adS space-time, either transparent or reflective boundary conditions can be imposed on a massless, conformally-coupled scalar field at time-like infinity \cite{Avis:1977yn}.
There are two possible reflective boundary conditions, which effectively correspond to either Dirichlet or Neumann boundary conditions on the conformal scalar field.
Here we make the choice $c_{q}=2$, which corresponds to Dirichlet reflective boundary conditions on the scalar field as $\zeta \rightarrow \infty $.
The effect on the renormalized VP of choosing alternative boundary conditions will be explored elsewhere \cite{Morley} (see also \cite{Worden:2018} for the effect of boundary conditions on the renormalized VP on a BTZ black hole).

To find $q_{n\lambda }$, we rewrite the homogeneous version of the radial equation (\ref{eq:radialODE}) as a differential equation with independent variable $\zeta ^{-1}$, start our integration at a large value of $\zeta $, using the series (\ref{eq:qseries}) with $c_{q}=2$ to give initial conditions on $q_{n\lambda }$ and its first derivative, and then integrate inwards towards the event horizon.
For both $p_{n\lambda }(\zeta )$ and $q_{n\lambda }(\zeta )$, we use the Mathematica routine {\tt {NDSolve}} to integrate the ODE and store the radial functions and their derivatives on a grid of values of $\zeta =1 +10^{-2+3i/100}$ for integer $i\in [1,100]$. 
Our grid has a greater density of points near the event horizon $\zeta =1$, since we anticipate that the VP will be more quickly varying there.

The Wronskian of $p_{n\lambda }(\zeta )$ and $q_{n\lambda }(\zeta )$ enables the normalization constants $N_{n\lambda}$ to be determined via
\begin{equation}
 \left( p_{n\lambda } \frac{\rmd q_{n\lambda }}{\rmd \zeta } 
- q_{n\lambda } \frac{\rmd p_{n\lambda }}{\rmd \zeta } \right) 
= -\frac{N_{n\lambda }}{(\zeta ^{2}-1) h(\zeta )} .
\label{eq:wronskian}
\end{equation}
To check the accuracy of our solutions, we evaluate $N_{n\lambda }$ from our numerically computed $p_{n\lambda }(\zeta )$ and $q_{n\lambda }(\zeta )$ (and their derivatives) at each point on our grid, since this should be the same at all grid points for each $n$ and $\lambda $.
We find that the relative error
\begin{equation}
\frac{\max \left\{ N_{n\lambda } \right\} -\min \left\{ N_{n\lambda } \right\}}{\min \left\{ N_{n\lambda } \right\}}
\end{equation}
in our numerical values of $N_{n\lambda }$ for each $n$ and $\lambda $ considered is typically of the order of $10^{-47}$--$10^{-50}$.
The normalization constants $N_{n\lambda }$ are then combined with the radial functions $p_{n\lambda }(\zeta )$ and $q_{n\lambda }(\zeta )$ to give the radial Green's function $g_{n\lambda }(\zeta )$ (\ref{eq:radialg}). 
We will discuss the range of values of $n$ and $\lambda $ for which we find the radial functions in the next subsection.
 
\begin{figure}[h]
\begin{center}
\includegraphics[width=10cm]{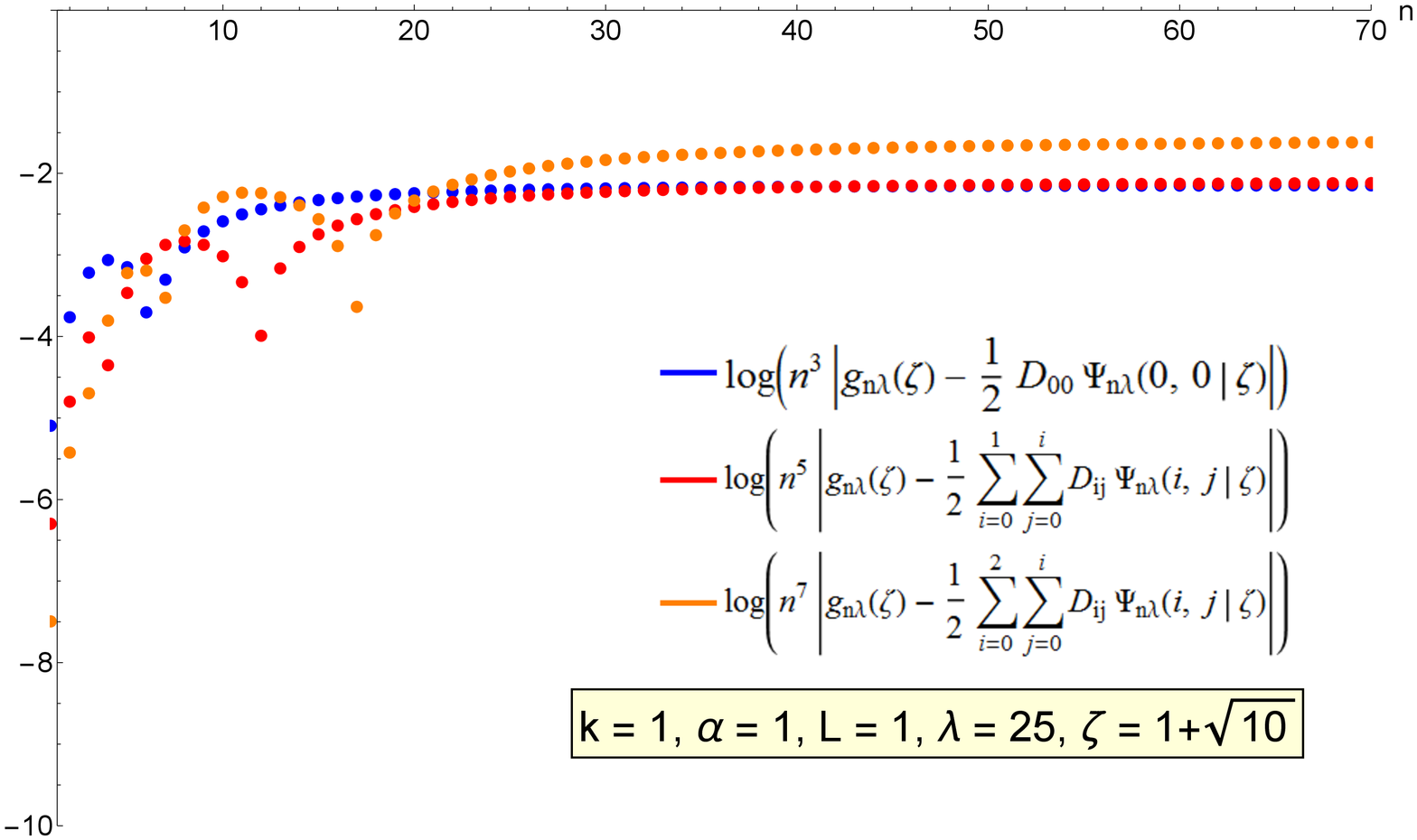}\\[0.2cm]
\includegraphics[width=10cm]{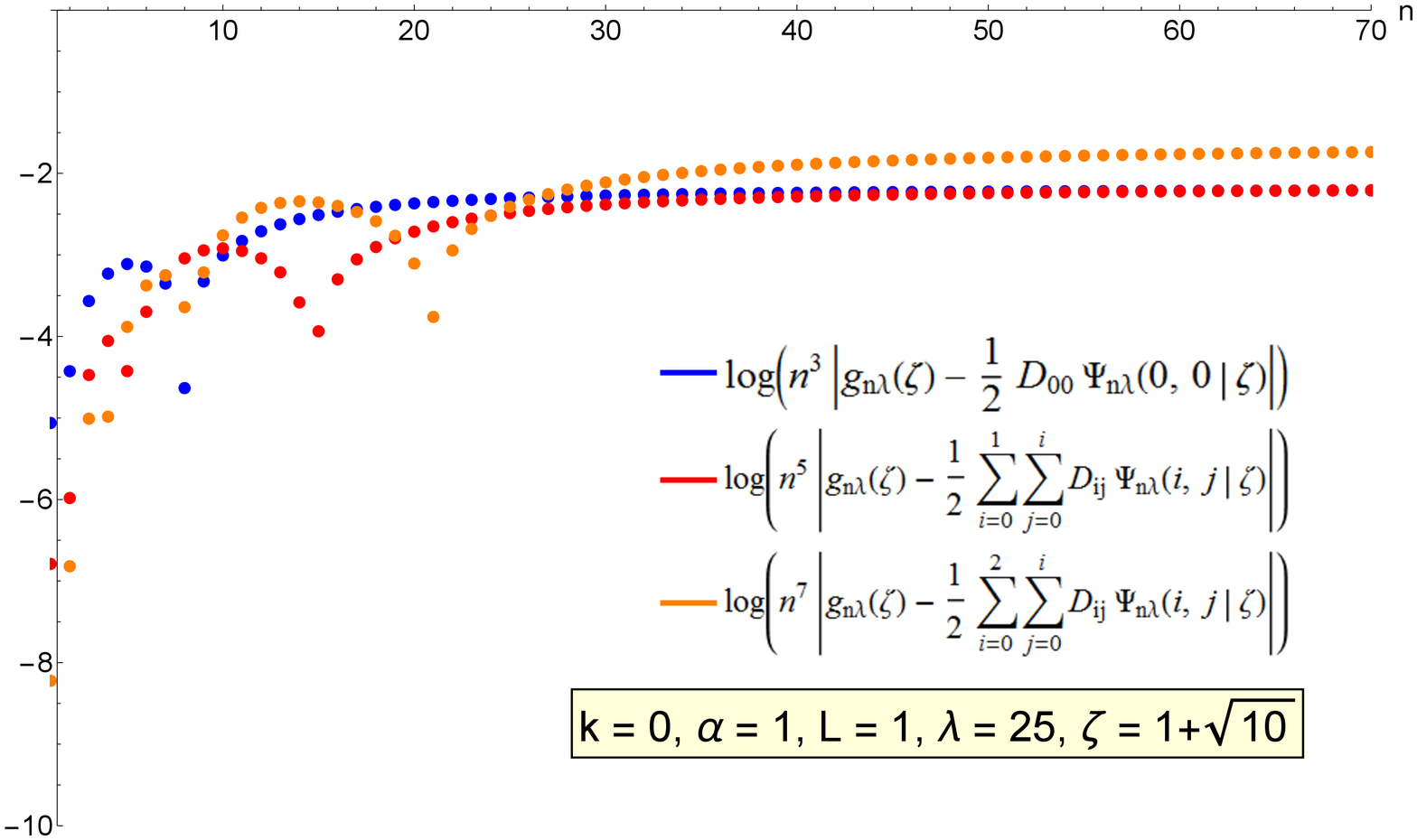}\\[0.2cm]
\includegraphics[width=10cm]{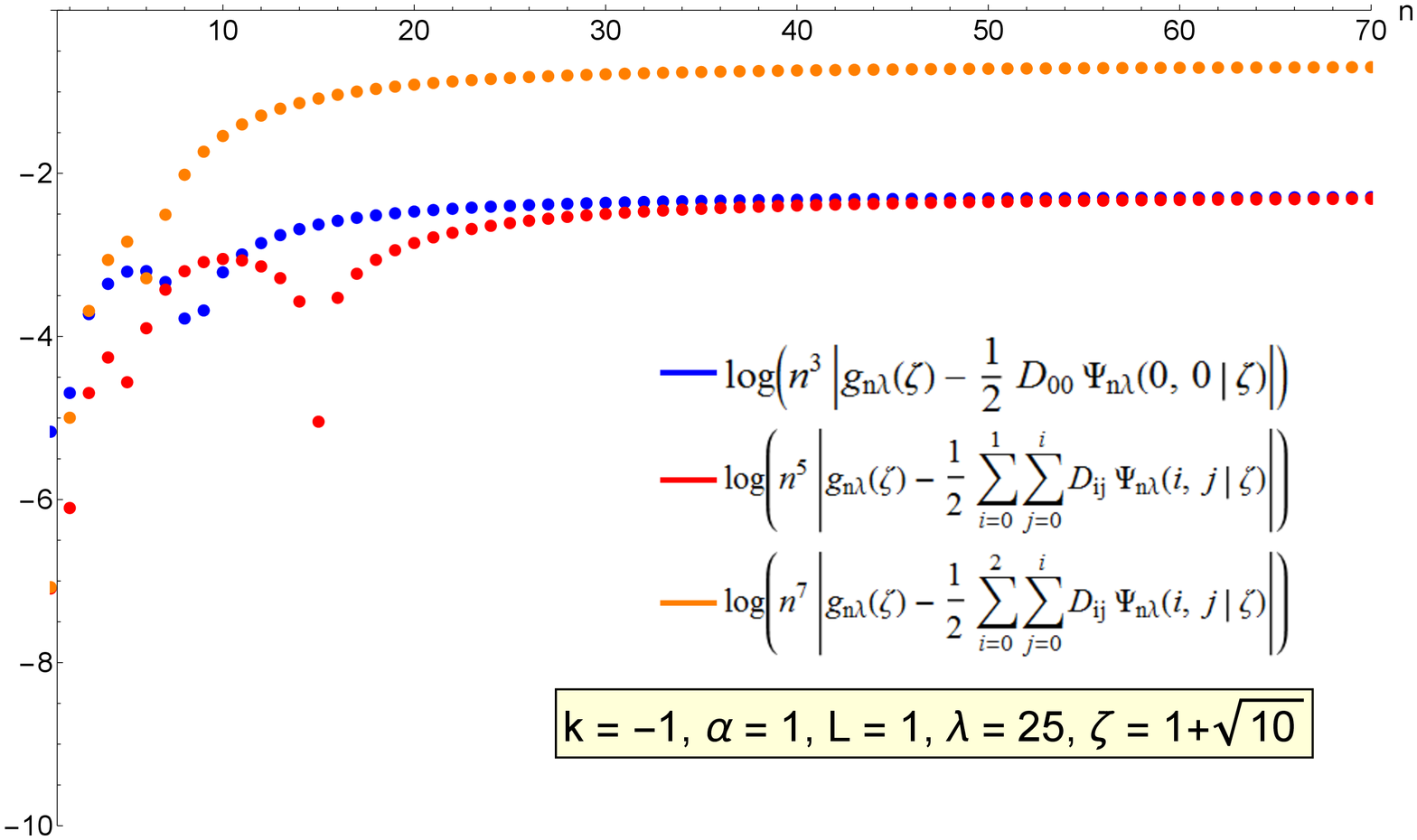}\\[0.2cm]
\end{center}
\caption{Log plots showing the convergence over $n$ in the mode sum expression (\ref{eq:VPfinal}). The blue lines show $\log \left|
\kappa g_{n\lambda}(\zeta) -\frac{1}{2}{\mathcal {D}}_{00}^{+}\Psi _{n\lambda }(0,0|r) \right| $, which is $\Or (n^{-3})$ for large $n$.
The red lines show $\log \left|
\kappa g_{n\lambda}(\zeta) - \frac{1}{2}\sum _{i=0}^{1}\sum _{j=0}^{i}{\mathcal {D}}_{ij}^{+}\Psi _{n\lambda }(i,j|r) \right| $, which is $\Or (n^{-5})$ for large $n$.
The orange lines show $\log \left|
\kappa g_{n\lambda}(\zeta) -\frac{1}{2} \sum _{i=0}^{2}\sum _{j=0}^{i}{\mathcal {D}}_{ij}^{+}\Psi _{n\lambda }(i,j|r) \right| $, which is $\Or (n^{-7})$ for large $n$.
The top plot is for $k=1$, the middle plot for $k=0$, and the bottom plot for $k=-1$. In each case we have set $\alpha =1$, $L=1$, $\lambda =25$ and $\zeta =1+1/{\sqrt{10} }$. Similar results are found for other values of the parameters. 
}
\label{fig:nconvergence}
\end{figure}

\subsection{Mode sums}
\label{sec:modesums}

Having found the radial functions $p_{n\lambda }$ and $q_{n\lambda }$,
the regularization parameters $\Psi _{n\lambda }(i,j|r)$ (\ref{eq:Psifinal}) are readily computed in Mathematica since we have analytic expressions for these quantities.
For each $i$, $j$, we simplify the derivatives in (\ref{eq:Psifinal}) using standard relations for the Legendre, Bessel and conical functions, which speeds up the computation. 
It remains therefore to combine the radial functions and regularization parameters in the mode sums (\ref{eq:VPfinal}).
It is more straightforward computationally to evaluate the sum over $n$ before finding the integral over $\lambda $.

The sums over $n$ in (\ref{eq:VPfinal}) converge extremely rapidly, as can be seen in Figure \ref{fig:nconvergence}.
Here, for each value of $k=1,0,-1$, we have chosen the space-time parameters to be $\alpha =1$, $L=1$, and plotted the absolute values of the quantities 
$\kappa g_{n\lambda}(\zeta )
-\frac{1}{2}\mathcal{D}_{00}^{(+)}(r)\Psi_{n\lambda}(0,0|r)$ (blue curves),
$\kappa g_{n\lambda}(\zeta )
-\frac{1}{2}\sum_{i=0}^{1}\sum_{j=0}^{i}\mathcal{D}_{ij}^{(+)}(r)\Psi_{n\lambda}(i,j|r)$ (red curves) and
$\kappa g_{n\lambda}(\zeta )
-\frac{1}{2}\sum_{i=0}^{2}\sum_{j=0}^{i}\mathcal{D}_{ij}^{(+)}(r)\Psi_{n\lambda}(i,j|r) $ (orange curves) for $\lambda =25$ and $\zeta =1 +1/{\sqrt {10}}$. 
If we subtract just the $i=0$ regularization parameters, the summand is $\Or (n^{-3})$ for large $n$ (see blue curves in Figure \ref{fig:nconvergence}).
Subtracting the $i=1$ regularization parameters as well as those for $i=0$, the convergence is quicker and the summand is $\Or (n^{-5})$ (red curves in Figure \ref{fig:nconvergence}). Finally, subtracting the $i=2$ regularization parameters as well gives a summand which is $\Or (n^{-7})$ and a very rapidly converging sum (orange curves in Figure \ref{fig:nconvergence}).
We find similar results for other values of $\lambda $ and $\zeta $ and the space-time parameters $\alpha $ and $L$. For each $k$, the asymptotic values of $\left| \kappa g_{n\lambda}(\zeta )
-\frac{1}{2}\mathcal{D}_{00}^{(+)}(r)\Psi_{n\lambda}(0,0|r)\right| $ and $\left| \kappa g_{n\lambda}(\zeta )
-\frac{1}{2}\sum_{i=0}^{1}\sum_{j=0}^{i}\mathcal{D}_{ij}^{(+)}(r)\Psi_{n\lambda}(i,j|r)\right| $ for large $n$ are very similar, while those of $\left|\kappa g_{n\lambda}(\zeta )
-\frac{1}{2}\sum_{i=0}^{2}\sum_{j=0}^{i}\mathcal{D}_{ij}^{(+)}(r)\Psi_{n\lambda}(i,j|r) \right| $ are larger. This effect was also seen in \cite{Taylor:2016edd,Taylor:2017sux} and is more marked for $k=-1$ than for $k=0$ or $1$.
In our mode sums, we have used values of $n$ such that $\left| n \right| \le 70$. 
We have tested the error involved in truncating the mode sum at $\left| n \right| = 70$ by also evaluating the sums over $n$ with $\left| n \right| \le 60$. The relative error between these two mode sums is typically $\Or (10^{-9})$ or smaller.

\begin{figure}[h]
\begin{center}
\includegraphics[width=7.5cm]{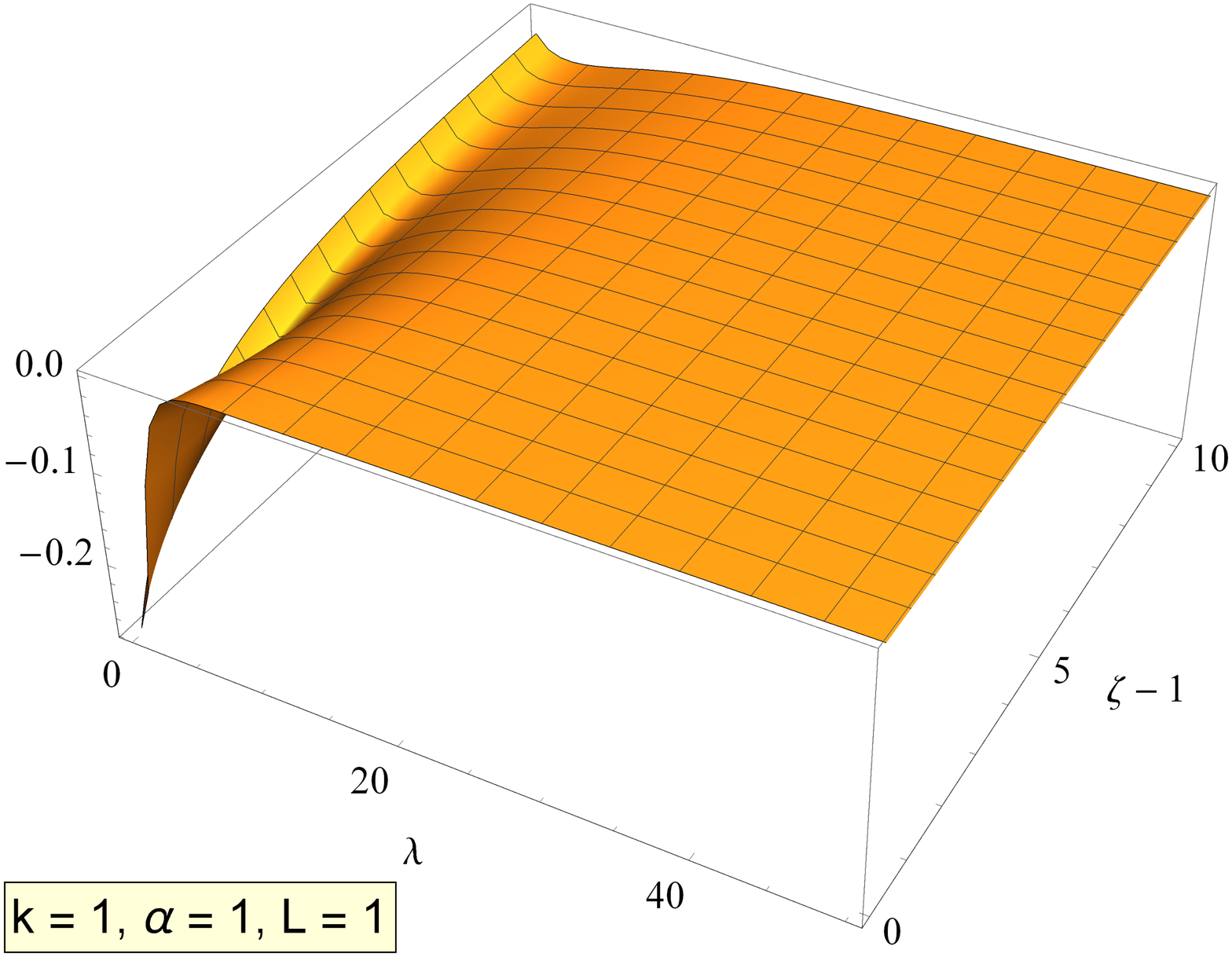}\\[0.2cm]
\includegraphics[width=7.5cm]{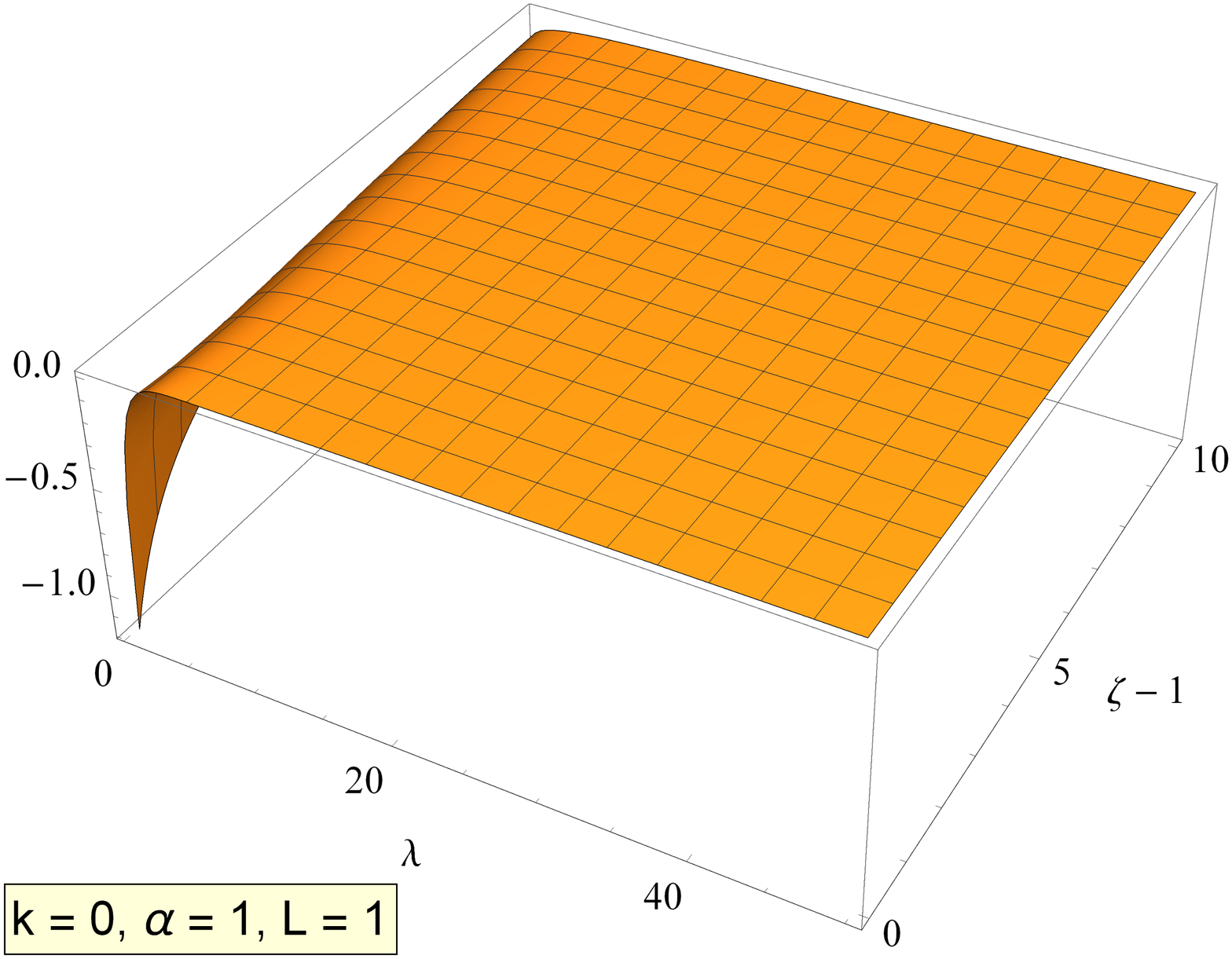}\\[0.2cm]
\includegraphics[width=7.5cm]{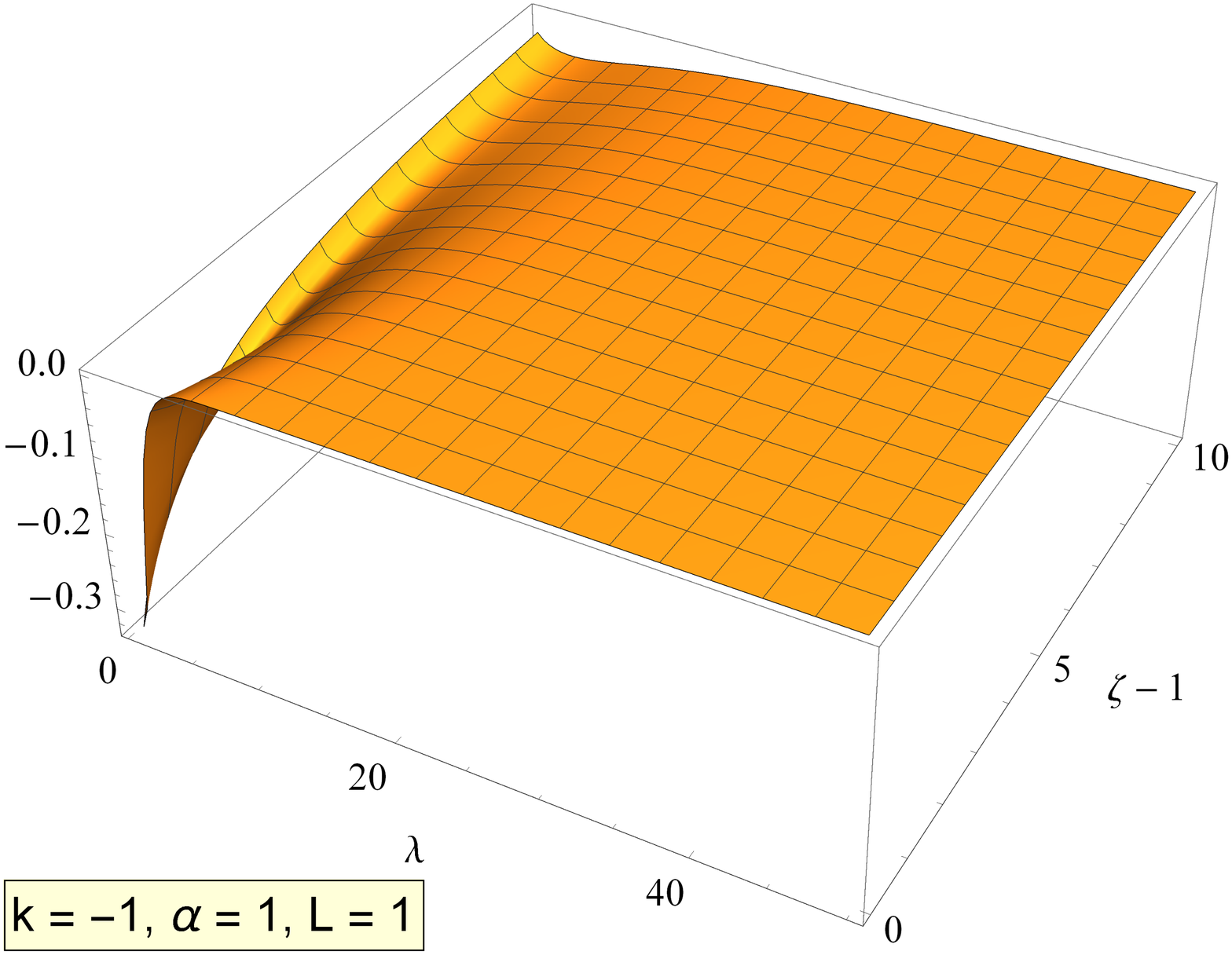}\\[0.2cm]
\end{center}
\caption{Plots showing the convergence over $\lambda $ in the mode sum expression (\ref{eq:VPfinal}). 
We plot the integrand/summand $\sum _{n}\mathcal{P}_{\lambda}^{(k)}(0)
\left[ \kappa g_{n\lambda}(\zeta )
-\frac{1}{2}\sum_{i=0}^{2}\sum_{j=0}^{i}\mathcal{D}_{ij}^{(+)}(r)\Psi_{n\lambda}(i,j|r)
\right]$ 
as a function of $\lambda $ and $\zeta $. 
The top plot is for $k=1$, the middle plot for $k=0$, and the bottom plot for $k=-1$. In each case we have set $\alpha =1$ and $L=1$.
Similar results are found for other values of the parameters. It can be seen that the integrand/summand converges rapidly to zero as $\lambda $ increases, but that this convergence is not uniform. For $k=\pm 1$, the convergence is more rapid for smaller values of $\zeta $, while for $k=0$ the convergence appears to be more rapid as $\zeta $ increases.}
\label{fig:lconvergence}
\end{figure}

Once the sums over $n$ have been computed, the next stage is to perform the sum/integral over $\lambda $ in (\ref{eq:VPfinal}). 
For $k=1$, this is a sum over $\lambda = 0,1,2,\ldots $, while for $k=0$ and $-1$, we have an integral over $\lambda \in [0,\infty )$.
First we check that the sum/integral is convergent for large $\lambda $.
In Figure \ref{fig:lconvergence} we plot the integrand/summand $\sum_{n}\mathcal{P}_{\lambda}^{(k)}(0)
\left[ \kappa g_{n\lambda}(\zeta )
-\frac{1}{2}\sum_{i=0}^{2}\sum_{j=0}^{i}\mathcal{D}_{ij}^{(+)}(r)\Psi_{n\lambda}(i,j|r)
\right]$ 
as a function of $\lambda $ and $\zeta $, for each value of $k$ and space-time parameters $\alpha =1$, $L=1$. 
We find similar results for other values of $\alpha $ and $L$.
For each value of $k$, it can be seen that the integrand/summand converges rapidly to zero as $\lambda $ increases. 
However, this convergence is not uniform in $\zeta $.
When $k=\pm 1$, the convergence is more rapid for smaller values of $\zeta $, while for $k=0$ the convergence is more rapid for larger values of $\zeta $.

For $k=1$, we have to compute a sum over $\lambda = 0,1,2,\ldots $.
We find that the summand is $\Or (\lambda ^{-5})$ for large $\lambda $, but, as observed in Figure \ref{fig:lconvergence}, this convergence is not uniform in $\zeta $.
Furthermore, we find that the convergence also depends on the value of $\alpha $.
In general, higher values of $\lambda $ are required for good accuracy in our final answers if we either increase $\zeta $ with $\alpha $ fixed, or increase $\alpha $ with $\zeta $ fixed. 
Our general approach for testing the accuracy of our sums is the same as that taken above for the sum over $n$, namely, we find the sums over $\lambda =0,1,\ldots ,\lambda _{\rm {max}}$ and $\lambda =0,1,\ldots ,\lambda _{{\rm {max}}}-10$ for some $\lambda _{\rm {max}}$ and find the relative error between the two values obtained.
When $\alpha =1$, with $\lambda _{\rm {max}}=70$, we find a relative error of $\Or (10^{-9})$ for values of $\zeta $ close to the horizon and $\Or (10^{-6})$ for $\zeta \sim 10$. 
For $\alpha =0.1$, a similar relative error is found near the horizon when $\lambda _{\rm {max}}=70$ and we again find small errors for larger values of $\zeta $ as well.
If we consider $\alpha =10$, then although $\lambda _{\rm {max}}=70$ is sufficient to give a small relative error of $\Or (10^{-10})$ near the horizon, we find that we require $\lambda _{\rm {max}}=130$ to obtain a final answer with reasonable accuracy for $\zeta \sim 3$.

When $k=0$ or $-1$, we have an integral over positive values of $\lambda $ rather than a sum.
To evaluate the integral, we find the modes on a grid of values of $\lambda $ up to $\lambda =\lambda _{\rm {max}}$, and then construct a cubic spline interpolating between the grid points in $\lambda $.
The interpolating function is then integrated over $\lambda $.
This process involves two sources of error. First,  there is the error due to the fact that the integral is truncated at $\lambda =\lambda _{\rm {max}}$ rather than being performed over all positive values of $\lambda $.
Second, approximating the integrand by an interpolating function between the grid points in $\lambda $ also introduces an error into the evaluation of the integral.

To estimate the first error, we first assume that the integrand for large $\lambda $ takes the power law form $\delta _{\lambda }\lambda ^{-1-\epsilon _{\lambda }}$. 
Here $\delta _{\lambda }$ and $\epsilon _{\lambda }$ are constants which are estimated by fitting this power law (using the Mathematica function {\tt {FindFit}}) to the values of the integrand computed numerically for $\lambda \in \left( 40,50 \right)$. 
The values of $\epsilon _{\lambda }$ are approximately equal to $4$,
so that the behaviour of the integrand for large $\lambda $ matches that of the summand in the $k=1$ case.
Integrating the power law gives an estimate $\delta _{\lambda }\lambda _{\rm {max}}^{-\epsilon _{\lambda }}/\epsilon _{\lambda }$ for the remainder of the integral. 
We find that the relative error in neglecting this part of the integral is $\Or (10^{-5})$ with $\lambda _{\rm {max}}=50$ for values of $\zeta $ close to the event horizon, and increases as $\zeta $ increases. 

The integrand varies more rapidly for small values of $\lambda $ than for larger $\lambda $. We therefore employ a grid of values of $\lambda $ with more points for smaller values of $\lambda $. 
In particular, we use $80$ equally spaced values of $\lambda $ for $\lambda \in (0,4]$ and a further $92$ equally spaced values of $\lambda $ for $\lambda \in (4,50]$.
To estimate the error resulting from our grid, we compare the values of the integral obtained using the above grid with that obtained from using a grid containing only half the number of values of $\lambda $.
The relative error between the two answers is $\Or (10^{-3})$ close to the event horizon and decreases as $\zeta $ increases.
We see that this error is by far the largest contribution to the total error in our computation of the renormalized VP.  
It would be possible to reduce this error by decreasing the separation of the points in our grid of $\lambda $ values, but this would considerably increase the computational time required to find the radial functions.

Once the sum/integral over $\lambda $ has been performed, the final, trivial, part of the computation of the renormalized VP is to subtract the term $-f'(r)/48\pi ^{2}r$ in (\ref{eq:VPfinal}).

\subsection{Results for the vacuum polarization}
\label{sec:results}

We now present our results for the renormalized VP (\ref{eq:VPfinal}) on a selection of topological black holes. 
The black hole metric (\ref{eq:metric}) depends on three parameters: $k$ (which governs the curvature of the event horizon), $\alpha $ (or, equivalently, the adS curvature length-scale $L$) and the black hole mass $M$.
We first consider fixed $k$ and black hole mass $M=0.5$ with varying $\alpha $, and then fix $\alpha $ and $L$ and compare the results for different values of $k$.
For large values of $\zeta $, it will be instructive to compare our results for the renormalized VP on the black hole space-times with the renormalized VP for a massless, conformally-coupled scalar field in the vacuum state on pure adS space-time, which is given by \cite{Kent:2014nya} 
\begin{equation}
\langle \hat{\phi}^{2} \rangle _{\rm {adS,vac}}= - \frac{1}{48\pi ^{2}L^{2}}= -\frac{\alpha ^{2}\left(4\alpha ^{2}+k \right) ^{2}}{48\pi ^{2}M^{2}}.
\label{eq:adSVP}
\end{equation}
The vacuum pure adS VP (\ref{eq:adSVP}) is always negative, and its magnitude decreases as the adS curvature length-scale $L$ increases or $\alpha $ decreases.
In all our plots below, dots denote values computed numerically, while the curves interpolate between these values.

\begin{figure}
\begin{center}
\includegraphics[width=10cm]{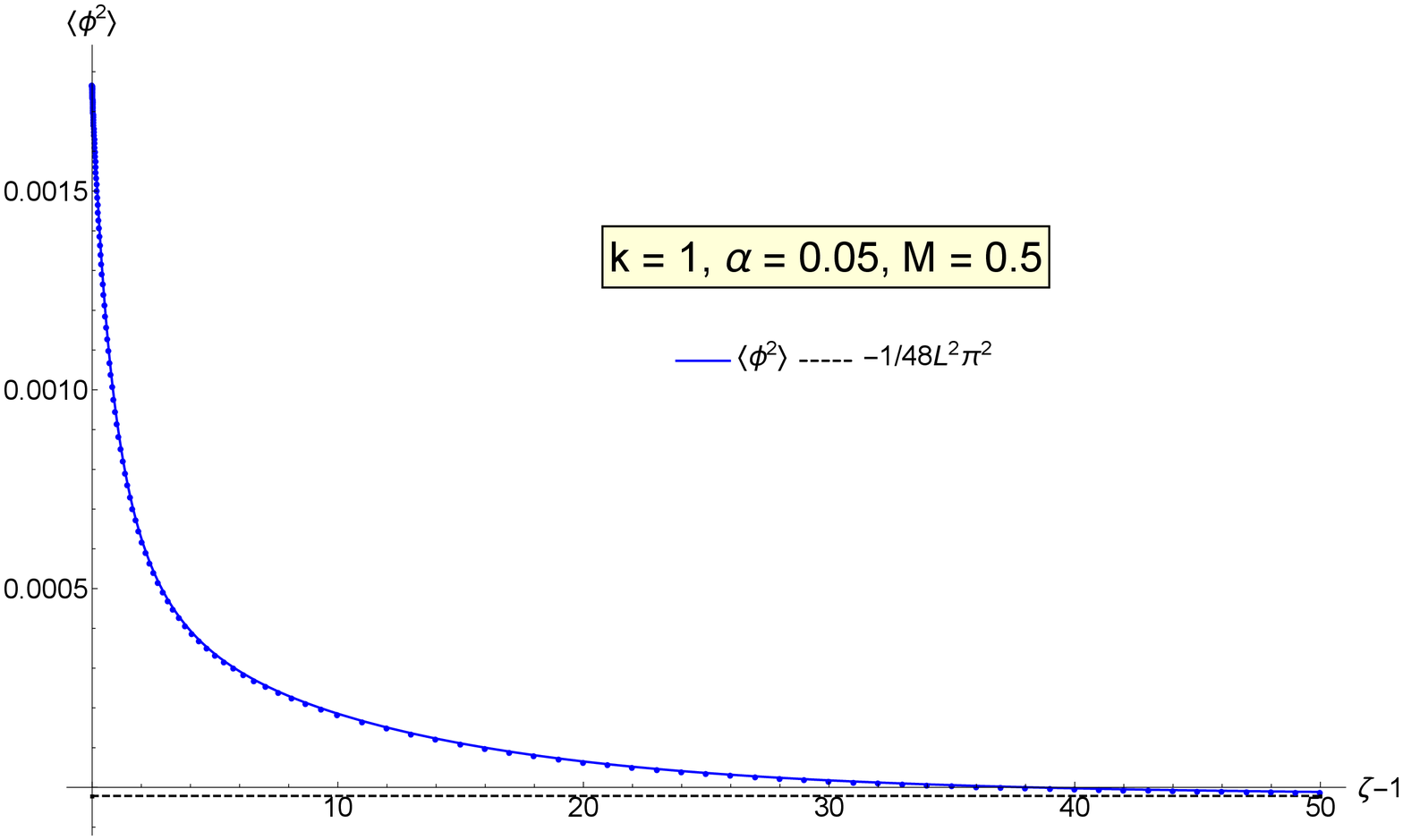}\\[0.2cm]
\includegraphics[width=10cm]{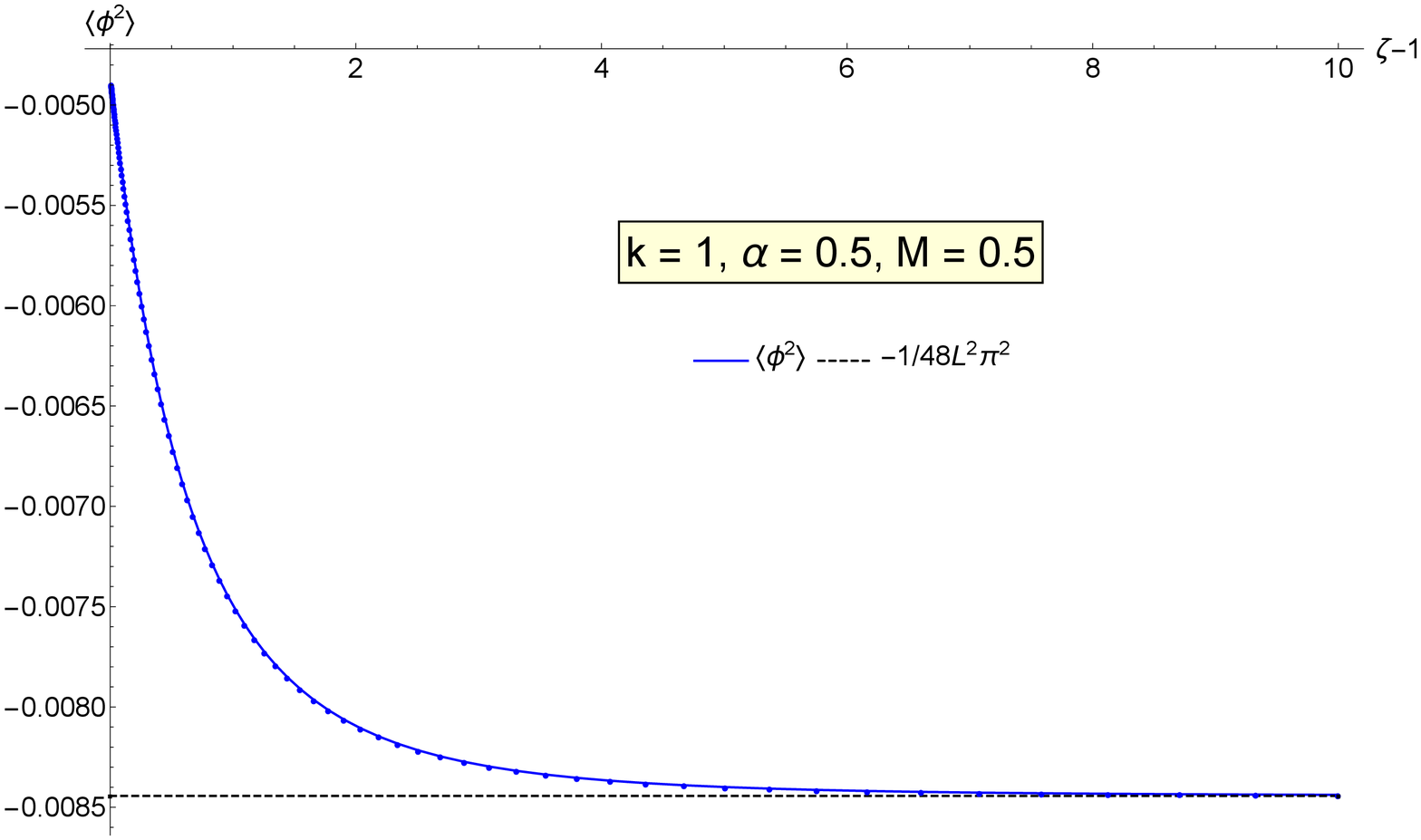}\\[0.2cm]
\includegraphics[width=10cm]{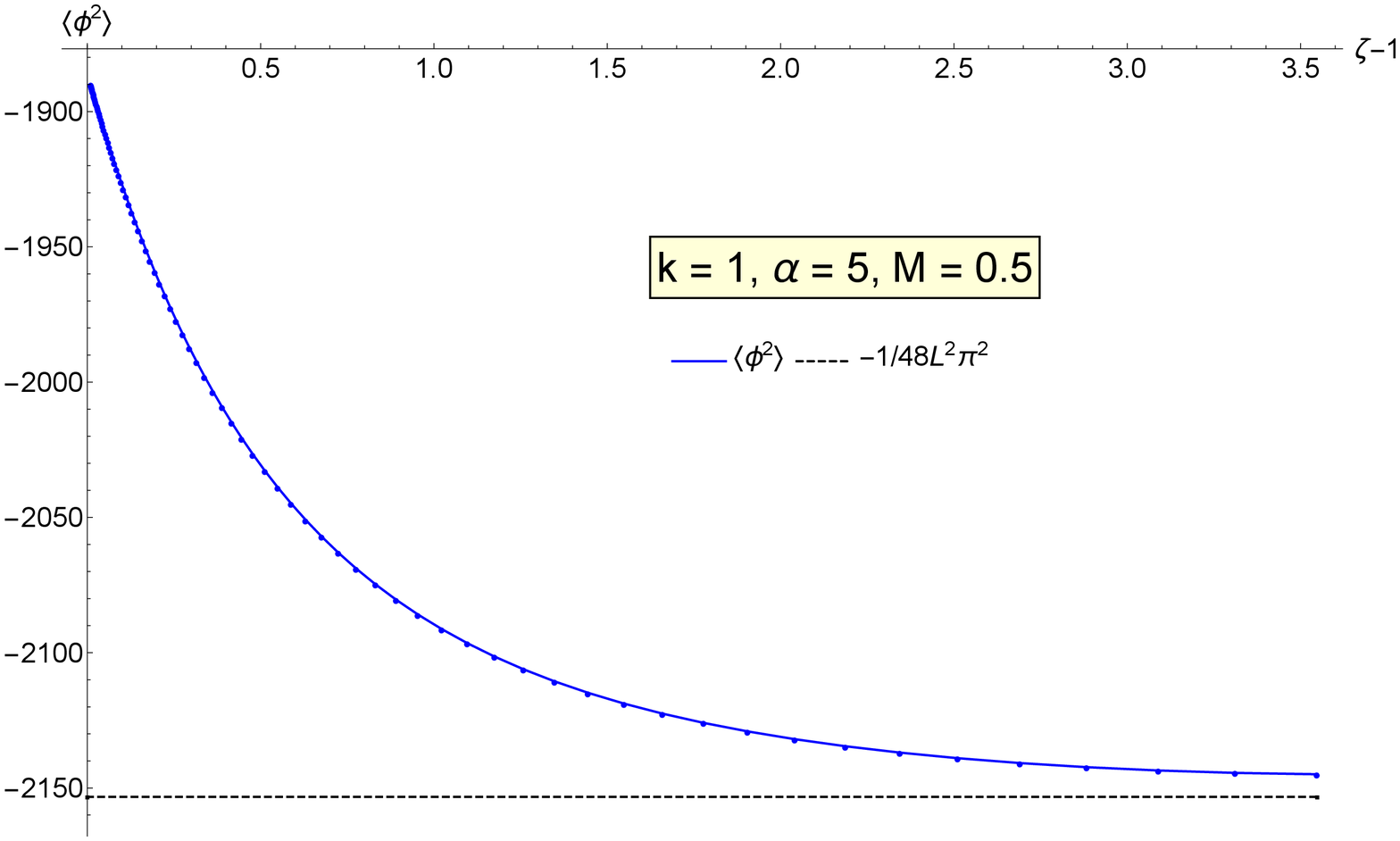}\\[0.2cm]
\end{center}
\caption{Renormalized VP (blue curves) for a massless, conformally-coupled scalar field on an asymptotically adS black hole with $k=1$, mass $M=0.5$ and three values of the space-time parameter $\alpha $: $\alpha =0.05$ (top), $\alpha = 0.5$ (middle) and $\alpha = 5$ (bottom). The renormalized VP is shown as a function of the dimensionless radial coordinate $\zeta $. The black dotted lines denote the renormalized VP for a massless, conformally-coupled scalar field in pure adS space-time (\ref{eq:adSVP}).}
\label{fig:k=1}
\end{figure}

\begin{figure}
\begin{center}
\includegraphics[width=10cm]{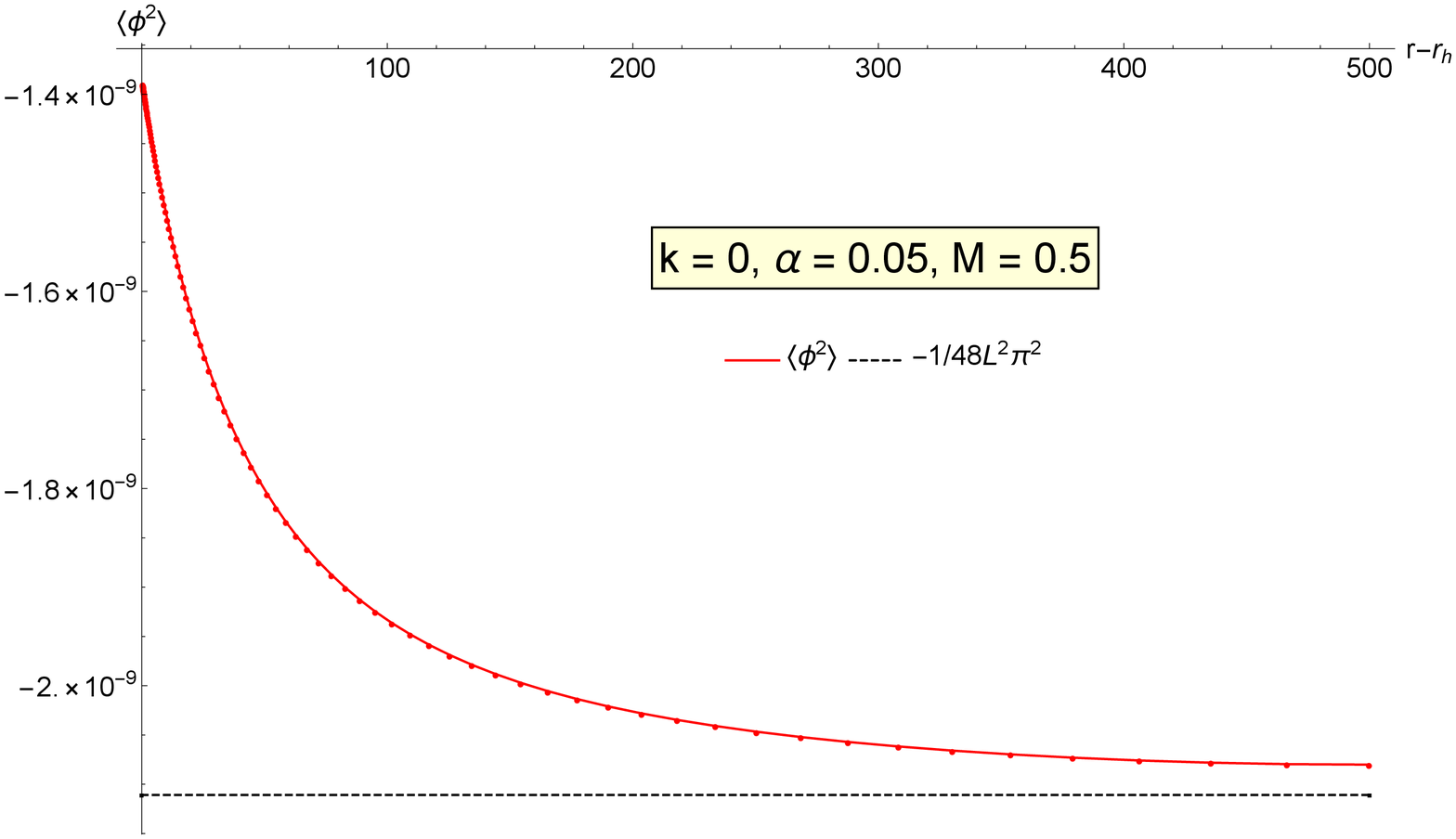}\\[0.2cm]
\includegraphics[width=10cm]{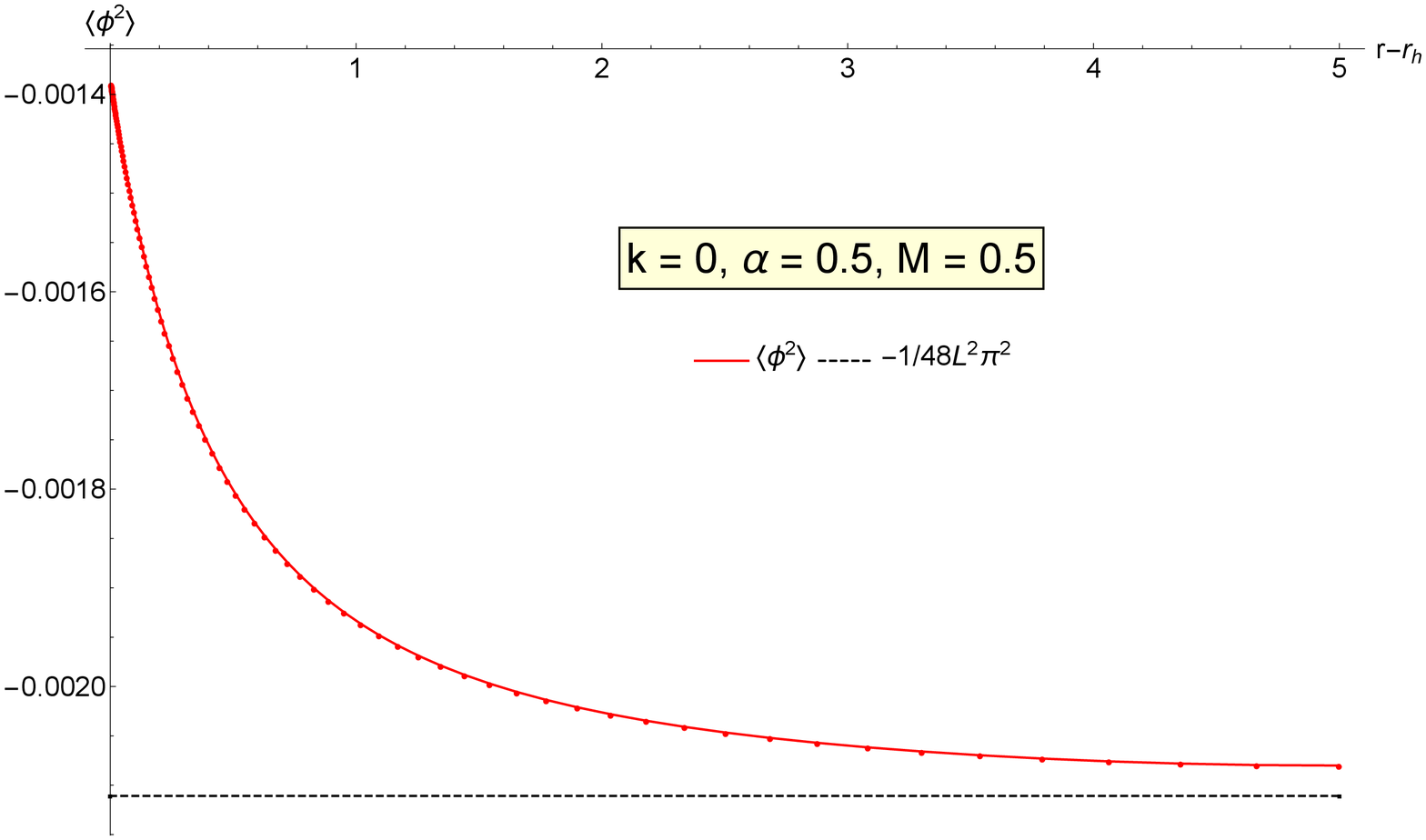}\\[0.2cm]
\includegraphics[width=10cm]{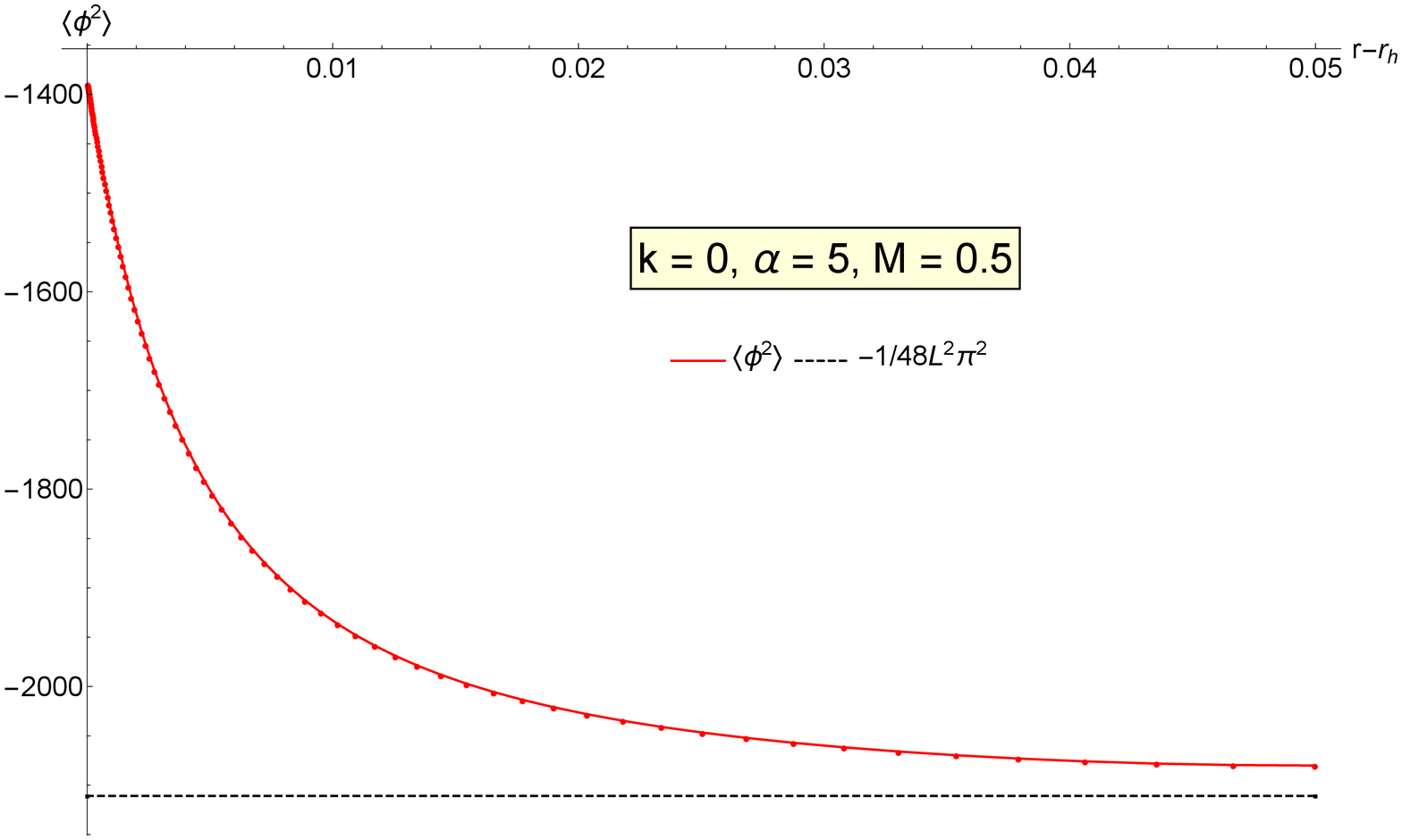}\\[0.2cm]
\end{center}
\caption{Renormalized VP (red curves) for a massless, conformally-coupled scalar field on an asymptotically adS black hole with $k=0$, mass $M=0.5$ and three values of the space-time parameter $\alpha $: $\alpha =0.05$ (top), $\alpha = 0.5$ (middle) and $\alpha = 5$ (bottom). The results for $\alpha =0.05$ and $\alpha =5$ are obtained from those for $\alpha =0.5$ by applying the scaling (\ref{eq:VPscaling2}). This scaling leaves the dimensionless radial coordinate $\zeta $ unchanged and thus the renormalized VP is shown as a function of the dimensionful radial coordinate $r$. The black dotted lines denote the renormalized VP for a massless, conformally-coupled scalar field in pure adS space-time (\ref{eq:adSVP}).}
\label{fig:k=0}
\end{figure}

\begin{figure}
\begin{center}
\includegraphics[width=10cm]{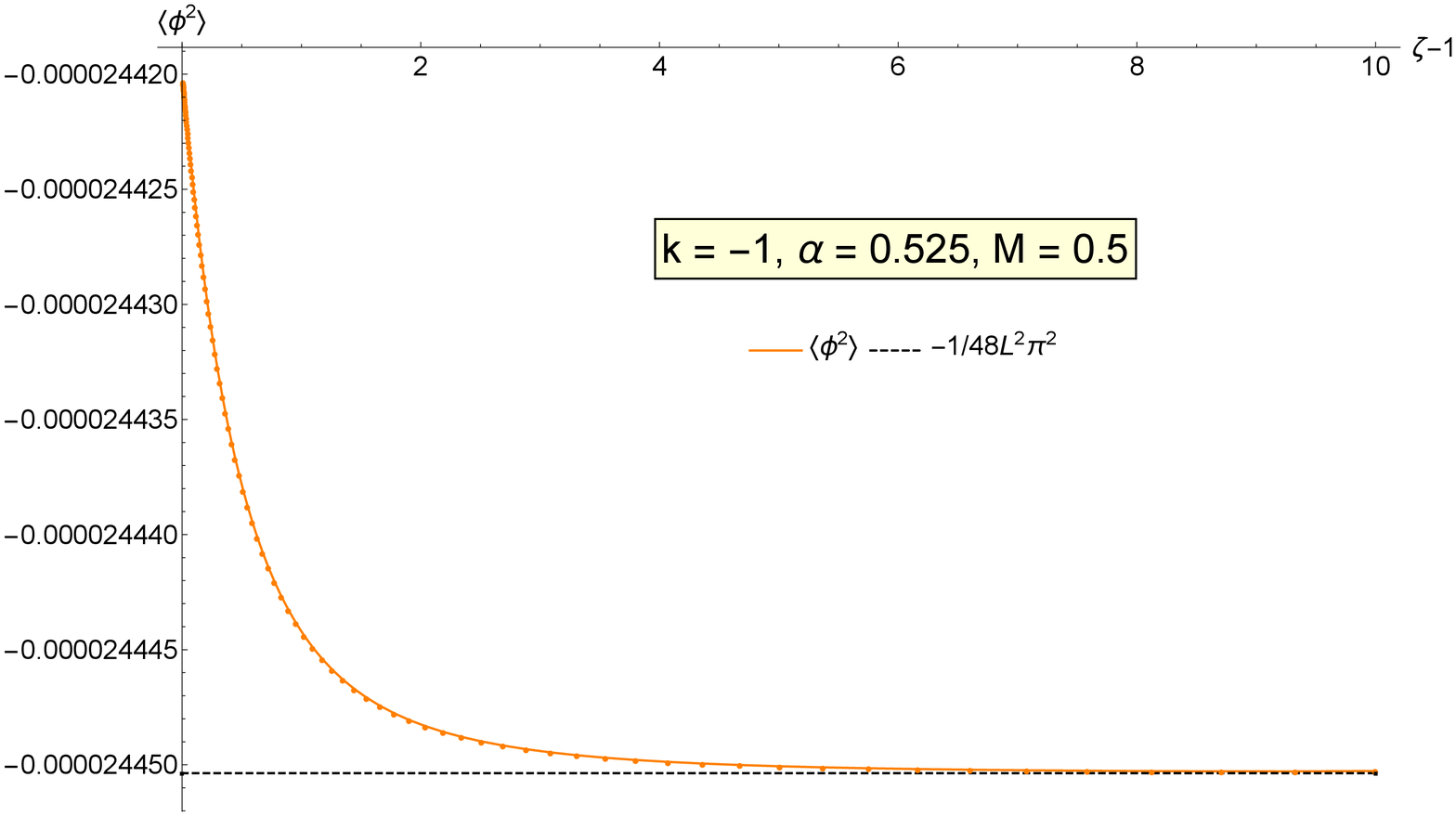}\\[0.2cm]
\includegraphics[width=10cm]{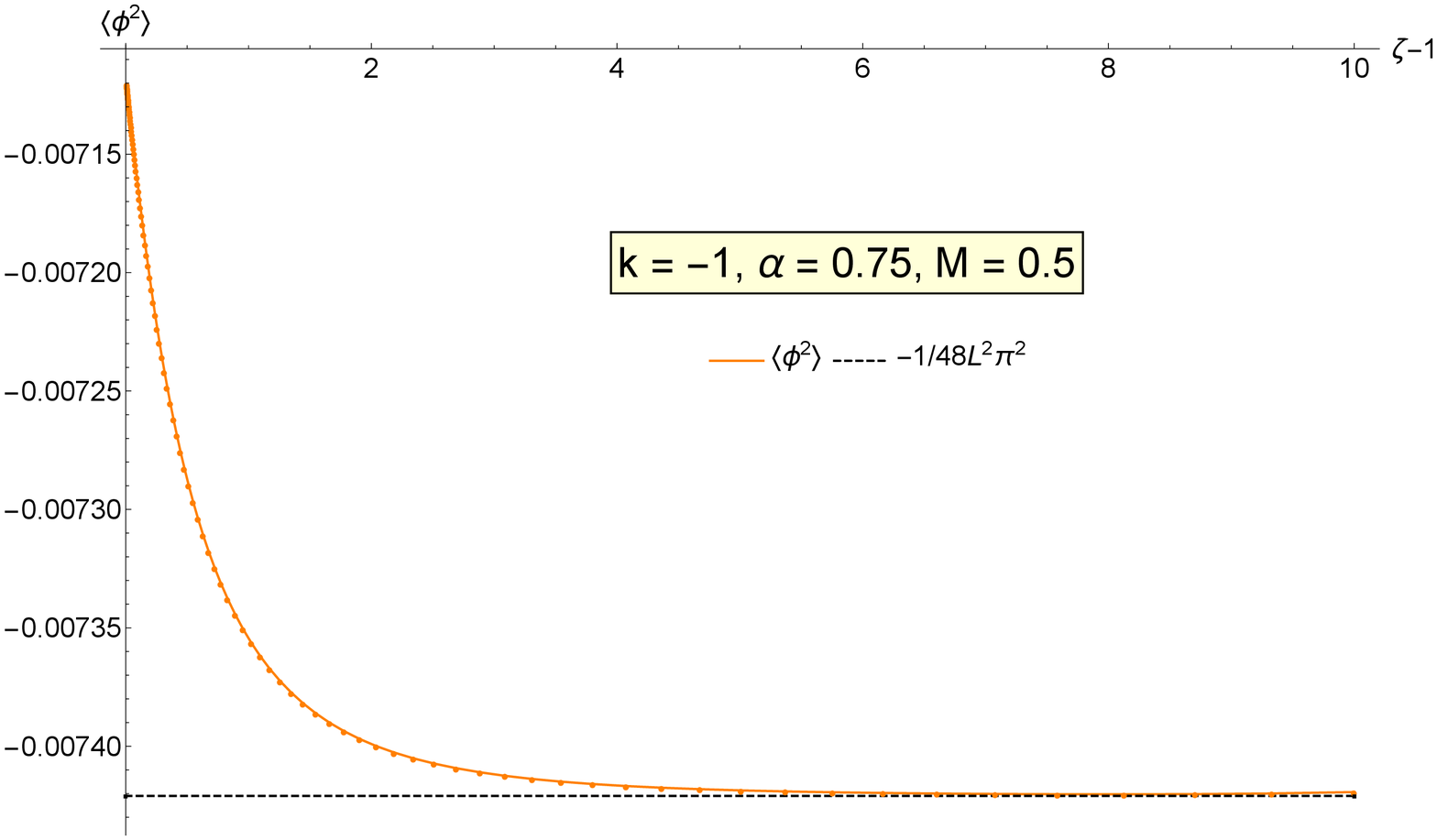}\\[0.2cm]
\includegraphics[width=10cm]{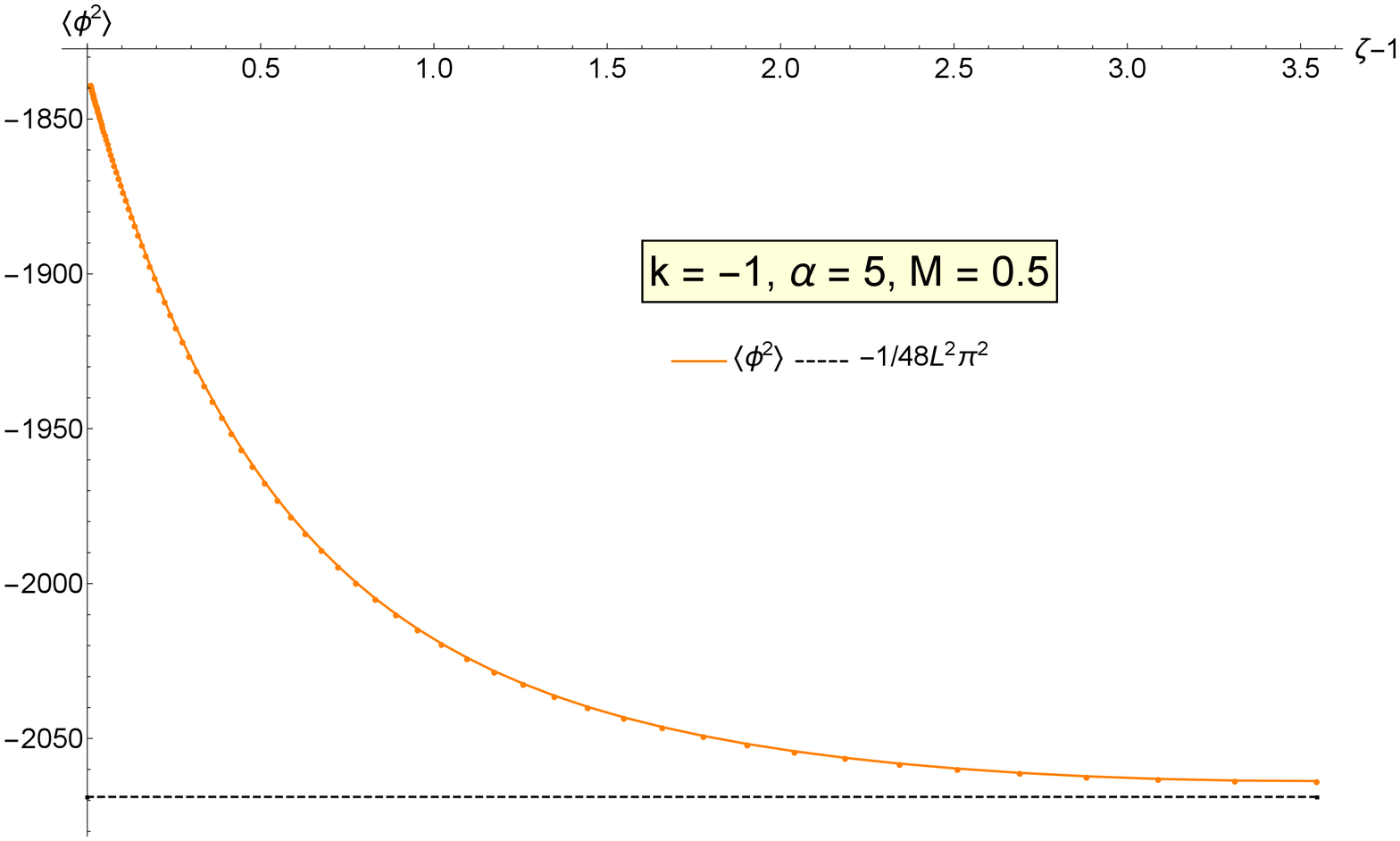}\\[0.2cm]
\end{center}
\caption{Renormalized VP (orange curves) for a massless, conformally-coupled scalar field on an asymptotically adS black hole with $k=-1$, mass $M=0.5$ and three values of the space-time parameter $\alpha $: $\alpha =0.525$ (top), $\alpha = 0.75$ (middle) and $\alpha = 5$ (bottom). The renormalized VP is shown as a function of the dimensionless radial coordinate $\zeta $. The black dotted lines denote the renormalized VP for a massless, conformally-coupled scalar field in pure adS space-time (\ref{eq:adSVP}).}
\label{fig:k=-1}
\end{figure}

We begin by considering spherical black holes with $k=1$. 
The renormalized VP in this case has been previously computed in \cite{Flachi:2008sr} using the AHS method, for fixed $L$ and various values of the black hole mass $M$. 
When $M=5$ and $L=1$, we find excellent agreement between the results in \cite{Flachi:2008sr} and ours obtained using the extended coordinate method.
In Figure \ref{fig:k=1} we plot the renormalized VP for black holes with $k=1$ and fixed mass $M=0.5$, with $\alpha = 0.05$, $0.5$ and $5$, plotted as a function of the dimensionless radial coordinate $\zeta $.
In each case we also show the value (\ref{eq:adSVP}) of the renormalized VP for a massless, conformally-coupled, scalar field on pure adS space-time.
For all values of $\alpha $, the black hole renormalized VP approaches that on pure adS as $\zeta \rightarrow \infty $, with the VP converging to the pure adS value more quickly for larger values of $\alpha $.
This is because, with fixed black hole mass $M$, increasing $\alpha $ corresponds to reducing the adS curvature length-scale $L$ (\ref{eq:alpha}), which means that the adS boundary is effectively closer to the event horizon.  
Alternatively, for large $\alpha $ and small $L$, the black hole is large compared with the adS curvature length-scale. 

For larger values of $\alpha $, the renormalized VP is negative everywhere on and outside the event horizon, but for small values of $\alpha $, the renormalized VP is positive in a region close to the event horizon.
For all values of $\alpha $, we find that the renormalized VP is monotonically decreasing from its value on the horizon to the asymptotic value (\ref{eq:adSVP}).
The magnitude of the renormalized VP both on and outside the horizon increases as $\alpha $ increases.  

We now turn to the topological black holes whose renormalized VP has not previously been considered in the literature.
Our results for planar black holes with $k=0$ are shown in Figure \ref{fig:k=0}.
In this case the black hole metric possesses two scaling symmetries (\ref{eq:scaling1}, \ref{eq:scaling2}), both of which leave the dimensionless radial coordinate $\zeta $ unchanged.
Applying the first scaling (\ref{eq:scaling1}), the parameter $\alpha $ is unchanged, the radial functions $p_{n\lambda }(\zeta )$ and $q_{n\lambda }(\zeta )$ are unchanged, as are the normalization constants $N_{n\lambda }$ (\ref{eq:wronskian}), while the radial Green's function $g_{n\lambda }$, the surface gravity $\kappa $, the coefficients ${\mathcal {D}}_{ij}$ and the regularization parameters $\Psi _{n\lambda }(i,j|r)$ scale as follows:
\begin{eqnarray}
g_{n\lambda }& \rightarrow  &\rho ^{-1}g_{n\lambda }, \qquad \kappa \rightarrow \rho ^{-1}\kappa, 
\nonumber \\
{\mathcal {D}}_{ij} & \rightarrow & \rho ^{-2i} {\mathcal {D}}_{ij}, \qquad
\Psi _{n\lambda }(i,j|r) \rightarrow \rho ^{2i-2} \Psi _{n\lambda }(i,j|r).
\end{eqnarray}
This means that the renormalized VP (\ref{eq:VPfinal}) scales as
\begin{equation}
\langle \hat{\phi}^{2} \rangle_{\mathrm{HH}} \rightarrow
\rho ^{-2}\langle \hat{\phi}^{2} \rangle_{\mathrm{HH}} .
\label{eq:VPscaling1}
\end{equation}
Under the second scaling (\ref{eq:scaling2}), 
the radial functions and $\alpha $ are unchanged, and we have
\begin{eqnarray}
N_{n\lambda } & \rightarrow \rho ^{2}N_{n\lambda },  \qquad  g_{n\lambda } \rightarrow \rho ^{-3}g_{n\lambda }, \qquad \kappa \rightarrow \rho \kappa, 
\nonumber \\ 
{\mathcal {D}}_{ij} & \rightarrow  \rho ^{2i+2j} {\mathcal {D}}_{ij}, \qquad
\Psi _{n\lambda }(i,j|r) \rightarrow \rho ^{-2i-2j-2} \Psi _{n\lambda }(i,j|r), 
\end{eqnarray}
which means that the renormalized VP is unchanged.
We can use a combination of these two scalings to scale $\alpha $ and $M$ independently.
Writing the renormalized VP as a function of $r$, $\alpha $ and the black hole mass $M$, and applying the scaling (\ref{eq:scaling1}) with parameter $\rho _{1}\rho _{2}^{-3}$, followed by the scaling (\ref{eq:scaling2}) with parameter $\rho _{2}$, we find
\begin{equation}
\langle \hat{\phi}^{2} \rangle_{\mathrm{HH}} (r, \alpha , M)=
\rho _{1}^{2}\rho _{2}^{-6}\langle \hat{\phi}^{2} \rangle_{\mathrm{HH}} (\rho _{1}\rho _{2}^{-2}r, \rho _{2}\alpha , \rho _{1}M).
\label{eq:VPscaling2}
\end{equation}
To compare with our results for $k=0$ with those for $k=1$ above and $k=-1$ below, we fix the black hole $M=0.5$. 
We compute the renormalized VP for $\alpha = 0.5$ and then use the scaling property (\ref{eq:VPscaling2}) to find the corresponding results for other values of $\alpha $.
Since the dimensionless radial coordinate $\zeta $ is unchanged by the scaling, we present our results for the renormalized VP as a function of the dimensionful radial coordinate $r$.

Unlike the situation for $k=1$, for planar black holes with $k=0$ we find that the renormalized VP is always negative at the event horizon (this would also be the case if we varied the black hole mass $M$, since the VP for different $M$ would be related to that for $M=0.5$ by the scaling relation (\ref{eq:VPscaling1})).
As in the $k=1$ case, the renormalized VP is monotonically decreasing as $r$ increases, and converges to the pure adS VP (\ref{eq:adSVP}) for large $r$.
The magnitude of the renormalized VP increases as $\alpha $ increases. 
The effect of the scaling on the radial coordinate $r$ means that the renormalized VP approaches its asymptotic values for large $r$ more slowly as $\alpha $ decreases, although since the scaling leaves $\zeta $ unchanged, the rate of convergence in $\zeta $ is unaffected by changing $\alpha $.
In this case, the scaling symmetries of the metric mean that the concept of a ``large'' or ``small'' black hole does not make sense. 

When $k=-1$ and the event horizon has constant negative curvature, the range of possible values of $\alpha $ for a fixed black hole mass $M$ is restricted by the definition (\ref{eq:alpha}).
In particular, when $M=0.5$, we find from considering the real root of (\ref{eq:alpha}) that $\alpha >0.5$. 
In Figure \ref{fig:k=-1} we plot the renormalized VP for $\alpha =0.525$, $0.75$ and $5$ as a function of the dimensionless radial coordinate $\zeta $.
Our results are qualitatively similar to those in the $k=1$ and $0$ cases, with the renormalized VP monotonically decreasing from its value on the event horizon as $\zeta $ increases.
On the event horizon, the renormalized VP is negative for all values of $\alpha $ studied. 
The magnitude of the renormalized VP increases as $\alpha $ increases, as was the case for the other two values of $k$.
We also see from Figure \ref{fig:k=-1} that the renormalized VP converges to its asymptotic value (\ref{eq:adSVP}) at smaller $\zeta $ as $\alpha $ increases. 
This is to be expected since, as in the $k=1$ case, large values of $\alpha $ correspond to small adS curvature length-scale $L$ and large black holes.

\begin{figure}
\begin{center}
\includegraphics[width=11cm]{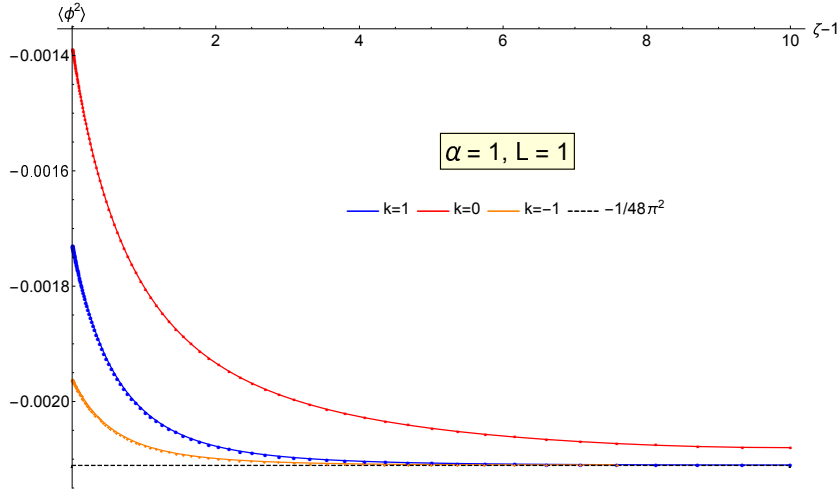}
\end{center}
\caption{Renormalized VP for a massless, conformally-coupled scalar field on an asymptotically adS black hole with $\alpha = 1$ and $L=1$. The renormalized VP is shown as a function of the dimensionless radial coordinate $\zeta $ for $k=1$ (blue curve), $k=0$ (red curve) and $k=-1$ (orange curve). The black dotted line denotes the renormalized VP for a massless, conformally-coupled scalar field in pure adS space-time (\ref{eq:adSVP}).}
\label{fig:allk}
\end{figure}

The plots in Figures \ref{fig:k=1}--\ref{fig:k=-1} show that the qualitative behaviour of the renormalized VP is very similar for all values of $k$, in particular the renormalized VP is monotonically decreasing from its value on the event horizon to the pure adS value (\ref{eq:adSVP}) for large $\zeta $.
In Figure \ref{fig:allk} we compare our results for different $k$ by fixing $\alpha = 1$ and the adS curvature length-scale $L=1$ (the black holes will then have different masses given by (\ref{eq:alpha}) for different $k$).
For these values of $\alpha $ and $L$, the renormalized VP is negative on the event horizon for all $k$. 
The magnitude of the renormalized VP on the horizon is of the same order of magnitude for all $k$, but it is larger for $k=0$  (red curve) and smaller for $k=-1$ (orange curve) compared with the $k=1$ case (blue curve).
The rate at which the renormalized VP approaches its asymptotic value (\ref{eq:adSVP}) also depends on $k$, with the $k=-1$ case converging most rapidly and the $k=0$ case converging the least rapidly.
For the $k=0$ case in particular, the renormalized VP is significantly different from its asymptotic value at comparatively large values of $\zeta $.

\section{Conclusions}
\label{sec:conc}

In this paper we have computed the renormalized VP for a massless, conformally-coupled scalar field on an asymptotically adS topological black hole space-time, for which the event horizon is a two-surface of constant curvature (either positive, negative, or zero).
The scalar field is in the Hartle-Hawking state \cite{Hartle:1976tp}, representing a black hole in thermal equilibrium with a heat bath at the Hawking temperature.

We have employed Hadamard renormalization and applied the extended coordinate method of \cite{Taylor:2016edd,Taylor:2017sux}, originally developed for spherically-symmetric black holes.
This method extends in an elegant way to topological black holes having an event horizon with either zero or negative curvature, and enables the Hadamard parametrix to be written as a mode sum.
As a result, the renormalized VP can be computed in a numerically efficient manner.
For spherically-symmetric black holes, the resulting expression for the renormalized VP involves a double sum over modes labelled by two integers.
One complication for topological black holes is that one of these sums is replaced by an integral over a continuous variable, which labels the eigenvalues of the Laplacian on the event horizon surface.
Finding the modes of the scalar field is a computationally intensive process, so we are only able to find these on a discrete grid of values of this continuous variable.
The main error in our calculation of the renormalized VP arises from integrating a function which interpolates between the points in this grid.
Nonetheless, we are able to present results for the renormalized VP which are accurate to at least three significant figures.

Our results show that the qualitative behaviour of the renormalized VP is similar for all event horizon topologies studied, being monotonically decreasing from its value on the event horizon to the (negative) vacuum pure adS value as the time-like adS boundary is approached.
This monotonically decreasing behaviour as the radial distance from the event horizon increases is the same as that observed for a massless conformally-coupled scalar field in the Hartle-Hawking state on an asymptotically flat Schwarzschild black hole \cite{Candelas:1984pg}.
In the asymptotically flat case, far from the black hole event horizon the renormalized VP approaches the constant value for a thermal state in Minkowski space-time with temperature equal to the black hole's Hawking temperature.
In view of this, it is perhaps surprising that in the asymptotically adS case the renormalized VP approaches the vacuum value far from the black hole. 
However, for thermal states of a quantum scalar field \cite{Ambrus:2018olh,Kent} on pure adS space-time the thermal radiation tends to ``clump'' away from the time-like adS boundary and the VP is not constant throughout the space-time.
Instead, near the boundary the VP for a thermal state on pure adS approaches that for the vacuum state, and we find similar behaviour for the thermal Hartle-Hawking state on the asymptotically adS black holes considered here.

For all the cases examined where the event horizon has zero or negative curvature, we find that the renormalized VP is negative everywhere on and outside the event horizon, but for small black holes with positive horizon curvature, the VP is positive in a region away from the adS boundary.
Small Schwarzschild-adS black holes with positive horizon curvature are thermodynamically unstable, while larger positive horizon curvature black holes are thermodynamically stable \cite{Hawking:1982dh}.
The black holes we consider with zero or negative horizon curvature are also thermodynamically stable \cite{Brill:1997mf}.
It may be that the sign of the renormalized VP at the horizon of an asymptotically adS black hole reflects the thermodynamic stability of the black hole (that is, if the VP is positive on the horizon, the black hole is thermodynamically unstable; if negative the black hole is thermodynamically stable), but testing this conjecture would involve significantly more computationally intensive calculations of the VP for different values of the space-time parameters than we have presented here.

In this paper we have considered the simplest nontrivial expectation value for a quantum scalar field, namely the VP, and found that the topology of the event horizon has interesting effects, both on the rate at which the VP converges to the vacuum adS value far from the horizon, and on the magnitude of the VP at the horizon.  
Here we have restricted our attention to a massless, conformally-coupled scalar field, and an open question remains whether these effects of the event horizon topology on the VP extend to the massive case or more general couplings to the space-time curvature. 
It would also be interesting to see the effect of the event horizon topology on the back-reaction of the quantum field on the space-time geometry. 
This requires a computation of the RSET, which is absent from the existing literature on asymptotically adS black holes even in the case where the event horizon has positive curvature.
Either of these extensions to the work presented here would involve finding a representation of the tail part of the Hadamard parametrix, which can be ignored when calculating the renormalized VP for a massless, conformally-coupled scalar field.
The extended coordinate method can be used to find a mode-sum representation for the tail \cite{Taylor:2017sux} when the event horizon has positive curvature.
Extending this method to horizons with zero or negative curvature is likely to be rather involved, so we leave this for future work. 

\ack
T.M.~thanks the School of Mathematics and Statistics at the University of Sheffield for the provision of a studentship supporting this work. 
The work of E.W.~is supported by the Lancaster-Manchester-Sheffield Consortium for Fundamental Physics under STFC grant ST/P000800/1 and partially supported by the H2020-MSCA-RISE-2017 Grant No.~FunFiCO-777740. 
E.W.~thanks Dublin City University for hospitality while this work was in progress.


\section*{References}

\begin{thebibliography}{99}

\bibitem{DeWitt:1975ys}
DeWitt B S 1975 {\it {Phys.~Rept.}} {\bf {19}} 295--357

\bibitem{Christensen:1976vb}
Christensen S M 1976 \PR D {\bf {14}} 2490--2501

\bibitem{Christensen:1978yd}
Christensen S M 1978 \PR D {\bf {17}} 946--963

\bibitem{Wald:1977up}
Wald R M  1977 {\it {Commun.~Math.~Phys.}} {\bf {54}} 1-19

\bibitem{Wald:1995yp}
Wald R M 1995
{\it {Quantum field theory in curved space-time and black hole
  thermodynamics}} 
(Chicago: University of Chicago Press)

\bibitem{Decanini:2005eg}
Decanini Y and Folacci A 2008 \PR D {\bf {78}} 044025 

\bibitem{Ottewill:2012mq}
Ottewill A C and Taylor P 2012 \PR D {\bf {86}} 104067

\bibitem{Kent:2014nya}
Kent  C and Winstanley E 2015 \PR D {\bf {91}} 044044

\bibitem{Fawcett:1983dk}
Fawcett M S 1983 {\it {Commun.~Math.~Phys.}} {\bf {89}} 103--115 

\bibitem{Howard:1984qp}
Howard K W  and Candelas P 1984 \PRL {\bf {53}} 403--406

\bibitem{Howard:1985yg}
Howard K W 1984 \PR D {\bf {30}} 2532--2547

\bibitem{Jensen:1992mv}
Jensen B P, McLaughlin J G and Ottewill A C 1992 \PR D {\bf {45}} 3002--3005

\bibitem{Ottewill:2010hr}
Ottewill A C and Taylor P 2011 \CQG {\bf {28}} 015007

\bibitem{Elster:1984hu}
Elster T 1984 \CQG {\bf {1}} 43--54

\bibitem{Groves:2002mh}
Groves P B, Anderson P R and Carlson E D 2002
\PR D {\bf {66}} 124017

\bibitem{Jensen:1988rh}
Jensen B P and Ottewill A C 1989 \PR D {\bf{39}} 1130--1138

\bibitem{Jensen:1995qv}
Jensen B P,   McLaughlin J G and Ottewill A C 1995 \PR D {\bf {51}} 5676--5697

\bibitem{Anderson:1993if}
Anderson P R, Hiscock W A and Samuel D A 1993  \PRL {\bf {70}} 1739--1742

\bibitem{Anderson:1994hg}
Anderson P R, Hiscock W A and Samuel D A 1995 \PR D {\bf {51}} 4337--4358

\bibitem{Levi:2016quh}
Levi A and Ori A 2016 \PRL {\bf {117}} 231101

\bibitem{Levi:2016exv}
Levi A, Eilon E, Ori A and van~de Meent M 2017 \PRL {\bf {118}} 141102

\bibitem{Levi:2015eea}
Levi A and Ori A 2015 \PR D {\bf {91}} 104028

\bibitem{Candelas:1980zt}
Candelas P 1980 \PR D {\bf {21}} 2185--2202

\bibitem{Candelas:1984pg}
Candelas P and Howard K W 1984 \PR D {\bf {29}} 1618--1625

\bibitem{Anderson:1989vg}
Anderson P R 1989 \PR D {\bf {39}} 3785--3788

\bibitem{Breen:2015hwa}
Breen C, Hewitt M, Ottewill A C and Winstanley E 2015 \PR D {\bf {92}} 084039

\bibitem{Levi:2016esr}
Levi A and Ori A 2016 \PR D {\bf {94}} 044054

\bibitem{Hartle:1976tp}
Hartle J B and Hawking S W 1976 \PR D {\bf {13}} 2188--2203

\bibitem{Boulware:1975pe}
Boulware D G 1975 \PR D {\bf {12}} 350--367

\bibitem{Unruh:1976db}
Unruh W G 1976 \PR D {\bf {14}} 870--892

\bibitem{Winstanley:2007tf}
Winstanley E and Young P M  2008 \PR D {\bf {77}} 024088

\bibitem{Breen:2010ux}
Breen C and Ottewill A C 2010 \PR D {\bf {82}} 084019

\bibitem{Breen:2011af}
Breen C and Ottewill A C 2012 \PR D {\bf {85}} 064026

\bibitem{Breen:2011aa}
Breen C and Ottewill A C  2012 \PR D {\bf {85}} 084029

\bibitem{Ottewill:2010bq}
Ottewill A C and Taylor P 2010 \PR D {\bf {82}} 104013

\bibitem{Ferreira:2014ina}
Ferreira H R C and Louko J  2015 \PR D {\bf {91}} 024038

\bibitem{Aharony:1999ti}
Aharony O, Gubser S S, Maldacena J M, Ooguri H and  Oz Y 2000 {\it {Phys.~Rept.}} {\bf {323}} 183--386

\bibitem{Flachi:2008sr}
Flachi A and Tanaka T 2008 \PR D {\bf {78}} 064011

\bibitem{Quinta:2016eql}
Quinta G M, Flachi A and Lemos J P S 2016 \PR D {\bf {93}} 124073

\bibitem{DeBenedictis:1998be} 
DeBenedictis A 1999 {\it {Gen.~Rel.~Grav.}} {\bf {31}} 1549--1563

\bibitem{Banados:1992wn}
Banados M, Teitelboim C and Zanelli J 1992 \PRL {\bf {69}} 1849--1851

\bibitem{Banados:1992gq}
Banados M, Henneaux M, Teitelboim C and Zanelli J 1993 \PR D {\bf {48}} 1506--1525, Erratum 2013 \PR D {\bf {88}} 069902

\bibitem{Carlip:1995qv}
Carlip S 1995 \CQG {\bf {12}} 2853--2880

\bibitem{Steif:1993zv}
Steif A R 1994 \PR D {\bf {49}} 585--589

\bibitem{Lifschytz:1993eb}
Lifschytz G and Ortiz M 1994 \PR D {\bf {49}} 1929--1943

\bibitem{Shiraishi:1993nu}
Shiraishi K and Maki T 1994 \CQG {\bf {11}} 695--700

\bibitem{Shiraishi:1993ti}
Shiraishi K and Maki T 1994 \CQG {\bf {11}} 1687--1696

\bibitem{Binosi:1998yu}
Binosi D, Moretti V, Vanzo L and Zerbini S 1999 \PR D {\bf {59}} 104017

\bibitem{Worden:2018}
Winstanley E and Worden K, paper in preparation

\bibitem{Martinez:1996uv}
Martinez C and Zanelli J 1997 \PR D {\bf {55}} 3642--3646

\bibitem{Casals:2016ioo}
Casals M, Fabbri A, Martinez C and Zanelli J 2016 \PL B {\bf {760}} 244--248

\bibitem{Casals:2016odj}
Casals M, Fabbri A, Martinez C and Zanelli J 2017 \PRL {\bf {118}} 131102

\bibitem{Birmingham:1998nr}
Birmingham D 1999 \CQG {\bf {16}} 1197--1205

\bibitem{Brill:1997mf}
Brill D R, Louko J and Peldan P 1997 \PR D {\bf {56}} 3600--3610

\bibitem{Lemos:1994fn}
Lemos J P S 1995 \CQG {\bf {12}} 1081--1086

\bibitem{Lemos:1994xp}
Lemos J P S 1995 \PL B {\bf {353}} 46--51

\bibitem{Lemos:1995cm}
Lemos J P S and Zanchin V T 1996 \PR D {\bf {54}} 3840--3853

\bibitem{Vanzo:1997gw}
Vanzo L 1997 \PR D {\bf {56}} 6475--6483

\bibitem{Cai:1996eg}
Cai R-G and Zhang Y-Z 1996 \PR D {\bf {54}} 4891--4898

\bibitem{Mann:1996gj}
Mann R B 1997 \CQG {\bf {14}} L109--L114

\bibitem{Smith:1997wx}
Smith W L and Mann R B 1997 \PR D {\bf {56}} 4942--4947

\bibitem{Mann:1997zn}
Mann R B 1998 {\it {Nucl.~Phys.}} B {\bf {516}} 357--381 

\bibitem{Taylor:2016edd}
Taylor P and Breen C 2016 \PR D {\bf {94}} 125024

\bibitem{Taylor:2017sux}
Taylor P and Breen C 2017 \PR D {\bf {96}} 105020

\bibitem{Birrell:1982ix}
Birrell N D and Davies P C W 1984
 {\it {Quantum fields in curved space}} (Cambridge: Cambridge University Press)
 
\bibitem{Avis:1977yn}
Avis S J, Isham C J and Storey D 1978 \PR D {\bf 18} 3565--3576 

\bibitem{Morley}
Morley T and Winstanley E, paper in preparation

\bibitem{Ambrus:2018olh}
 Ambrus V E, Kent C and Winstanley E 2018
{\it {Int.~J.~Mod.~Phys.}} D {\bf {27}} 1843014
  
\bibitem{Kent}
Kent C and Winstanley E, paper in preparation

\bibitem{Hawking:1982dh}
Hawking S W and Page D N 1983 {\it {Commun.~Math.~Phys.}}  {\bf 87} 577--588


\end{thebibliography}
\end{document}